\documentclass[twocolumn,secnumarabic,amsmath,nobibnotes,superscriptaddress,showpacs,aps,10pt,prb]{revtex4}
\usepackage{bm}%
\usepackage{color}
\usepackage{graphicx}
\usepackage{multirow}
\usepackage{subfigure}

\begin{document}
\title{Influence of defects on ferroelectric and electrocaloric properties of  BaTiO$_3$}%
\author{Anna Gr\"unebohm}
\email[Electronic mail: ]{anna@thp.uni-due.de}
\affiliation{Faculty of Physics and Center for Nanointegration, CENIDE, University of Duisburg-Essen, 47048 Duisburg, Germany}
\author{Takeshi Nishimatsu}
\affiliation{Institute for Materials Research (IMR), Tohoku University, Sendai 980-8577, Japan}

\begin{abstract}
We report modifications of the ferroelectric and electrocaloric properties of BaTiO$_3$ by defects.
For this purpose, we have combined \textit{ab initio}-based molecular dynamics simulations with a simple model for defects.  
We find that different kinds of defects modify the ferroelectric transition temperatures and polarization, reduce the thermal hysteresis of the transition and are no obstacle for a large caloric response.  For a locally reduced polarization the ferroelectric transition temperature and the adiabatic response are slightly reduced. For polar defects an intriguing picture emerges. 
The transition temperature is increased by polar defects and the  temperature range of the large caloric response is broadened. Even more remarkable, we find an inverse caloric effect in a broad temperature range.
 \end{abstract} \maketitle

\section{Introduction}
As conventional cooling is
presently one of the main carbon dioxide sources, more advanced technologies are urgently needed.
 In the last years the electrocaloric effect (ECE) has come into focus as a promising new cooling
mechanism.\cite{Moya,Madhura, Bai,Beckman,Ozbolt,Ponomareva,Scott} 
 Although the change of temperature in a ferroelectric material due to a varying electrical field is known
for decades, giant caloric responses up to 12~K have been obtained only recently.\cite{Moya} The understanding of the
ECE is still quite unsatisfactory. Materials with a large and reversible adiabatic response in a proper
temperature range, and which are ecologically save and economically viable, have to be found.

So far little is known about possible limitations of the ECE due to defects. However, defects such
as missing oxygen atoms or impurities and dopants are common in ferroelectric materials.  
Even though a large ECE of 0.5~K/300~kV/cm has been found
for doped BTO in an industrially manufactured multilayer capacitor,\cite{KarNarayan}
 we are not aware of any systematic investigation on the influence of
doping and defects on the ECE. The present paper aims to fill the gap in the understanding and optimization of the ECE in the presence of defects.

Different experimental and theoretical studies deal with the modification of ferroelectrics by doping and defects. \cite{Hagemann,Maglione,KarNarayan,Maier,Neumann,Park,choi2,Erhart,Erhart2,Erhart3}  
Both, doping and defects, influence remanent polarization and polarization switching, which are important for the caloric response. 
Evidently, even the modification of the ferroelectric
phase diagram under such doping is still unclear.  For example, reduction, insensitivity, and increase of the
ferroelectric transition temperature in BaTiO$_3$ (BTO) due to doping have been found.\cite{Hagemann,Maglione,KarNarayan,Maier}
In the present paper we use molecular dynamics simulations employing an effective Hamiltonian  based on \textit{ab initio} parametrization to improve the 
 understanding of the ferroelectric phase diagram by defects. The same model has been used successfully in order to model ferroelectric phase diagrams and the ECE in literature.\cite{Lisenkov, Ponomareva, Beckman, Madhura, Cp}

In particular, experimental and theoretical investigations reveal the
important role of transition metal doping on fatigue behavior in ferroelectric
perovskites.\cite{Erhart,Erhart2,Erhart3,Neumann} It has been found that an internal bias field builds up in 
field polarized (poled) Cr doped BTO, which could be related to the alignment of defect dipoles with the overall
polarization.\cite{Neumann} In this context Erhart and co-workers performed detailed \textit{ab initio} based transition state theory
simulations on transition metal doped perovskites.\cite{Erhart,Erhart2,Erhart3} They found that polar O-vacancy dopant
complexes, which induce local dipoles, form immediately. It is most favorable if these dipoles align parallel to the overall polarization. However, their relaxation is rather slow and the defect dipoles thus give rise to aging and fatigue,
i.e.\ broadening and shift of the field hysteresis.

Also, these changes of the phase diagram may largely modify the caloric
response and its reversibility. 
For the first time, the change of operation range and magnitude of the
ECE with different kinds of defects is investigated in the present work.  Also the influence of different measuring protocols is revealed.
With measuring protocol we mean the influence of the sample history, i.e.\ poled or as-prepared samples, cooling-down or heating-up simulations and the use of  
unipolar and bipolar electrical fields.

We show that a reduced local polarization reduces T$_C$, polarization, and the ECE, while local defect dipoles with slow relaxation dynamics and large dipole moment increase T$_C$ and may even induce an inverse caloric response.
The paper is organized as follows. First, the computational methods and the model for defects are discussed in Sec.~\ref{sec:comp}. A more detailed discussion on the accuracy of the method and a summary of convergency tests can be found in Sec.~\ref{sec:append}.
In Section~\ref{sec:nonpolar} the influence of non-polar defects on the ferroelectric phase diagram and the ECE are discussed. 
Section \ref{sec:polar} deals with the influence of polar complex dipoles on
the phase diagram of BTO and the ECE for different defects and measuring protocols. Finally,  conclusions and outlook are given in Sec.~\ref{sec:sum}.

\section{Computational Model}
\label{sec:comp}
\begin{figure}
\centering
\subfigure{(a)}{\includegraphics[height=25mm]{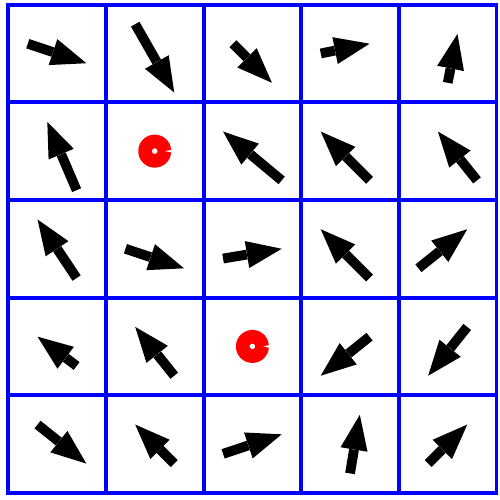}}
\subfigure{(b)}{\includegraphics[height=25mm]{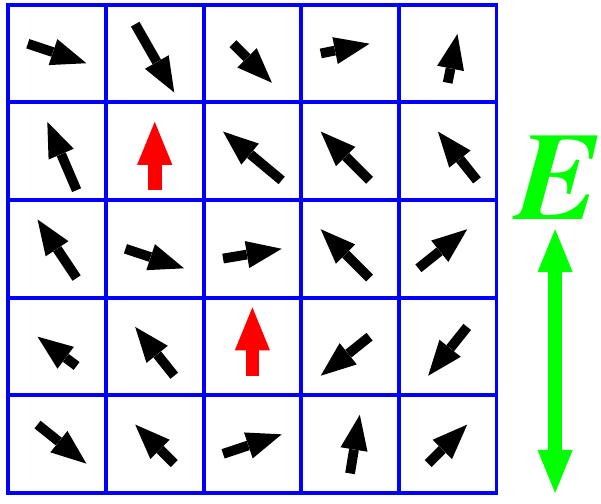}}
\caption{
  (Color online) Schematic sketch of the used defect model.
  (a) non-polar defects with locally frozen soft mode amplitude
  (b) fixed defect dipoles to which parallel or anti-parallel external electric field is applied.
}
\label{fig:model}
\end{figure}

Molecular dynamics (MD) simulations have been performed employing  the feram code
\url{http://loto.sourceforge.net/feram/}  developed by Nishimatsu \textit{ et al}.\cite{Feram1} based on an effective
Hamiltonian
\begin{multline}
  \label{eq:Effective:Hamiltonian}
  H^{\rm eff}
  = \frac{M^*_{\rm dipole}}{2} \sum_{\bm{R},\alpha}\dot{u}_\alpha^2(\bm{R})
  + \frac{M^*_{\rm acoustic}}{2}\sum_{\bm{R},\alpha}\dot{w}_\alpha^2(\bm{R})\\
  + V^{\rm self}(\{\bm{u}\})+V^{\rm dpl}(\{\bm{u}\})+V^{\rm short}(\{\bm{u}\})\\
  + V^{\rm elas,\,homo}(\eta_1,\dots\!,\eta_6)+V^{\rm elas,\,inho}(\{\bm{w}\})\\
  + V^{\rm coup,\,homo}(\{\bm{u}\}, \eta_1,\cdots\!,\eta_6)+V^{\rm coup,\,inho}(\{\bm{u}\}, \{\bm{w}\}),
\end{multline}
with $\eta_1,\dots,\eta_6$ the six components of homogeneous strain in Voigt notation.
$V^{\rm self}(\{\bm{u}\})$ is the self energy of the local mode,
$V^{\rm dpl}(\{\bm{u}\})$ is the long-ranged dipole-dipole interaction,
$V^{\rm short}(\{\bm{u}\})$ is the short-ranged interaction between local soft modes,
$V^{\rm elas,\,homo}(\eta_1,\dots,\eta_6)$ is the elastic energy from homogeneous strains,
$V^{\rm elas,\,inho}(\{\bm{w}\})$ is the elastic energy from inhomogeneous strains,
$V^{\rm coup,\,homo}(\{\bm{u}\}, \eta_1,\dots,\eta_6)$ is the coupling between the local soft modes and the homogeneous strain, and
$V^{\rm coup,\,inho}(\{\bm{u}\}, \{\bm{w}\})$ is the coupling between the soft modes and the inhomogeneous strains.

Instead of treating all atomic positions as degrees of freedom, the collective atomic displacements are coarse-grained by local soft mode
vectors $\bm{u}(\bm{R})$ and
local acoustic displacement vectors $\bm{w}(\bm{R})$
of each unit cell at $\bm{R}$ in a simulation supercell.
Therefore, the number of degrees of freedom per unit cell is reduced in a first step
from 5~atoms times the three cartesian directions ($= 15$)
to 2 three-dimensional vectors. 
$M^*_{\rm dipole}/2 \sum_{\bm{R},\alpha}\dot{u}_\alpha^2(\bm{R})$ and
$M^*_{\rm acoustic}/2\sum_{\bm{R},\alpha}\dot{w}_\alpha^2(\bm{R})$ are the
kinetic energies possessed by the local soft modes and
the local acoustic displacement vectors along with their effective masses of $M^*_{\rm dipole}$ and $M^*_{\rm acoustic}$.
In a second coarse graining step an internal optimization of  $E(\bm{w}(\bm{R}))$ is used reducing the degree of freedom to the three components of the soft mode $\bm{u}(\bm{R})$.
Details of this Hamiltonian are explained in Refs.~\onlinecite{King,Zhong,Feram1}.
The set of parameters for the effective Hamiltonian for BTO have been obtained by density functional theory simulation at $T=0$~K and are
listed in Ref.~\onlinecite{Nishimatsu}. 

Within the molecular dynamics (MD) simulations, periodic boundary conditions and a cell size of $96\times96\times96$ are used unlike otherwise stated. This corresponds to 884736 f.u.\ i.e.\ 4423680 atoms.
For cooling-down and heating-up simulations on a dense temperature grid of at most 5 K steps, the pre-converged configuration of previous steps is used as input for each temperature step and thus the equilibration time can be reduced to 40.000 fs.

The procedure of direct simulation of the ECE involves three steps:
first, we equilibrate the system in an external electrical field, E along [001], constrained to a constant number of particles ($N$), constant pressure ($P$), and constant temperature ($T$)  ($NPT$).
For this purpose we apply the Nos{\'e}-Poincar\'e thermostat.\cite{Nose} 
 Next, we change to adiabatic conditions keeping the number of particles $N$, the pressure ($P$), and the total energy ($E$) constant (micro canonical $NPE$ ensemble)   and let the ensemble evolve in time by the leapfrog method.
 Simultaneously, the field is ramped down 
 with a rate of 0.05~kV/cm/fs, see Sec.~\ref{sec:append} for a detailed discussion.
 The final state at the end of the constant temperature MD is used as the
initial state for the constant energy MD. 
 As last step we monitor the kinetic energy after 10.000 further equilibration steps. A time step of at most $\Delta t = 2$~fs is used in both ensembles.

In the Hamiltonian, only the three components of the soft mode vector are explicitly taken into account. Thus, the number of degrees of freedom is reduced from 15 to 3 which in turn reduces the specific heat. As a result the adiabatic response obtained within our model is overestimating the response of real BTO. However this overestimation is an intrinsic feature of the model and thus the qualitative trends for different kinds of defects are not affected. In addition, it has been shown that the leading error of this deviation can be corrected with a rescaling of the temperature by 15/3,\cite{Nishimatsu3,Beckman} see Sec.~6. In the following all results are rescaled correspondingly in order to get an estimate of the actual temperature change.

Defects are modeled by freezing the local soft mode on random positions to a certain value.
A local reduction of polarization by a defect is modeled for the limiting case of non-polar defects. Therefore, the local soft mode amplitude in some cells is frozen to zero, see Fig.~\ref{fig:model}~(a). 
Defect dipoles with slow relaxation are modeled by dipoles with fixed amplitude and direction, see Fig.~\ref{fig:model}~(b). We assume a ferroelectric coupling between defect dipoles and the surrounding BTO matrix. 
Thus, we investigate the influence of the modified dipole-dipole interaction by local changes of the polarization.
The used coarse graining procedure does not allow to model atomic relaxations near the defect. In addition, we neglect the direct coupling of defects to the external electrical field and perturbations of the ferroelectric matrix near the defect, e.g., by a modified volume or modified short range interactions.\\
For further details on the method and its excellent reliability for the modeling of the ferroelectric phase diagram,\cite{Zhong,Zhong3} the ECE, \cite{Lisenkov,Ponomareva,Beckman} or the influence of small distortions to the ideal system, e.g., due to  strain,\cite{Dieguez, Madhura} alloying,\cite{Lisenkov3} or finite-size effects\cite{Lai,Feram2} we refer to literature.

\section{{BaTiO$_3$} and non-polar defects}
\label{sec:nonpolar}
\begin{figure}
\includegraphics[height=0.35\textwidth,clip,trim=0cm 0.5cm 1cm 0.5cm]{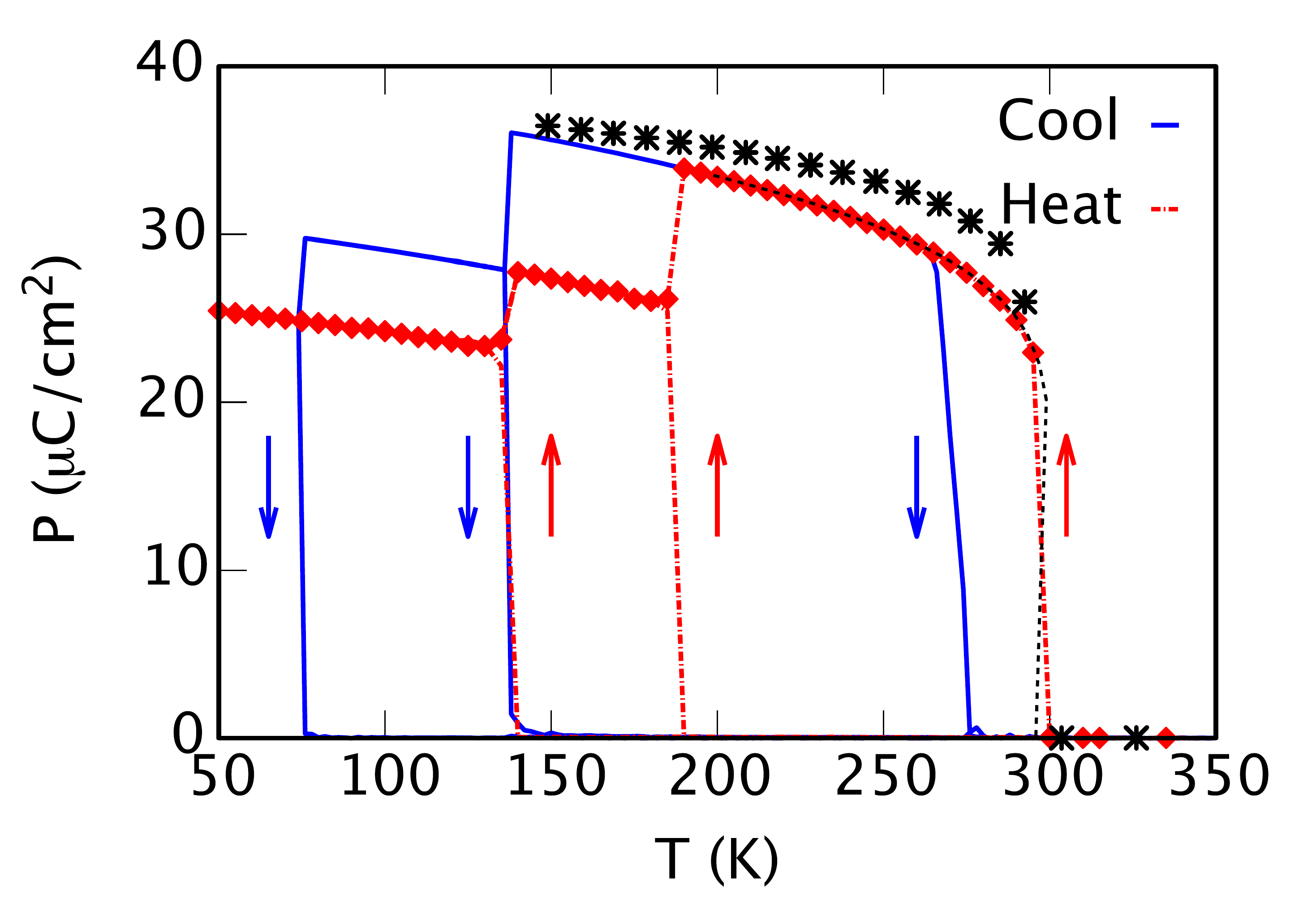}
\caption{(Color online) Polarization as function of temperature of BaTiO$_3$; Red: heating-up simulation; Blue: cooling-down simulation; Black: the sample has been equilibrated in 100~kV/cm and polarization and temperature have been recorded after field removal (ECE protocol). In all cases the polarization along [100],[010] and [001] is shown.\\
For the ECE protocol the data without pressure correction (dashed black lines) is opposed to results for an external pressure correction of $p=-0.005T$, treating the acoustic degrees of freedom explicitly, and shifted by 150 K to lower temperatures (black stars).
}
 \label{fig:Pol}
\end{figure}

The simulated polarization as function of temperature of BTO is illustrated in Fig.~\ref{fig:Pol}. At high temperatures BTO is in the paraelectric cubic perovskite structure. Under cooling, three
coupled structural-ferroelectric phase transitions exist, to the tetragonal, orthorhombic, and rhombic phases, 
with spontaneous polarization along [100], [110], and [111], respectively.  We find transition temperatures of 275~K, 150~K,
and 70~K for cooling-down simulations in qualitative agreement with the experimentally found phase sequence. 
In addition, spontaneous polarization and strain of all phases as well as the first order character of all transitions are as in experiment.\cite{Kumar}  The polarization jumps by about 30~$\mu C/cm^2$ at
$T_C$ and a pronounced  thermal hysteresis is present between cooling-down and heating-up because we neglect nucleation sites such as defects, surfaces, or grain boundaries, see Fig.~\ref{fig:Pol}. 
Experimentally, a thermal hysteresis of about 10~K has been found for BTO ceramics.\cite{Kanata}
\begin{figure}
  \centering
  \includegraphics[width=0.35\textwidth,clip,angle=270,trim=0.5cm 1cm 0cm 0cm]{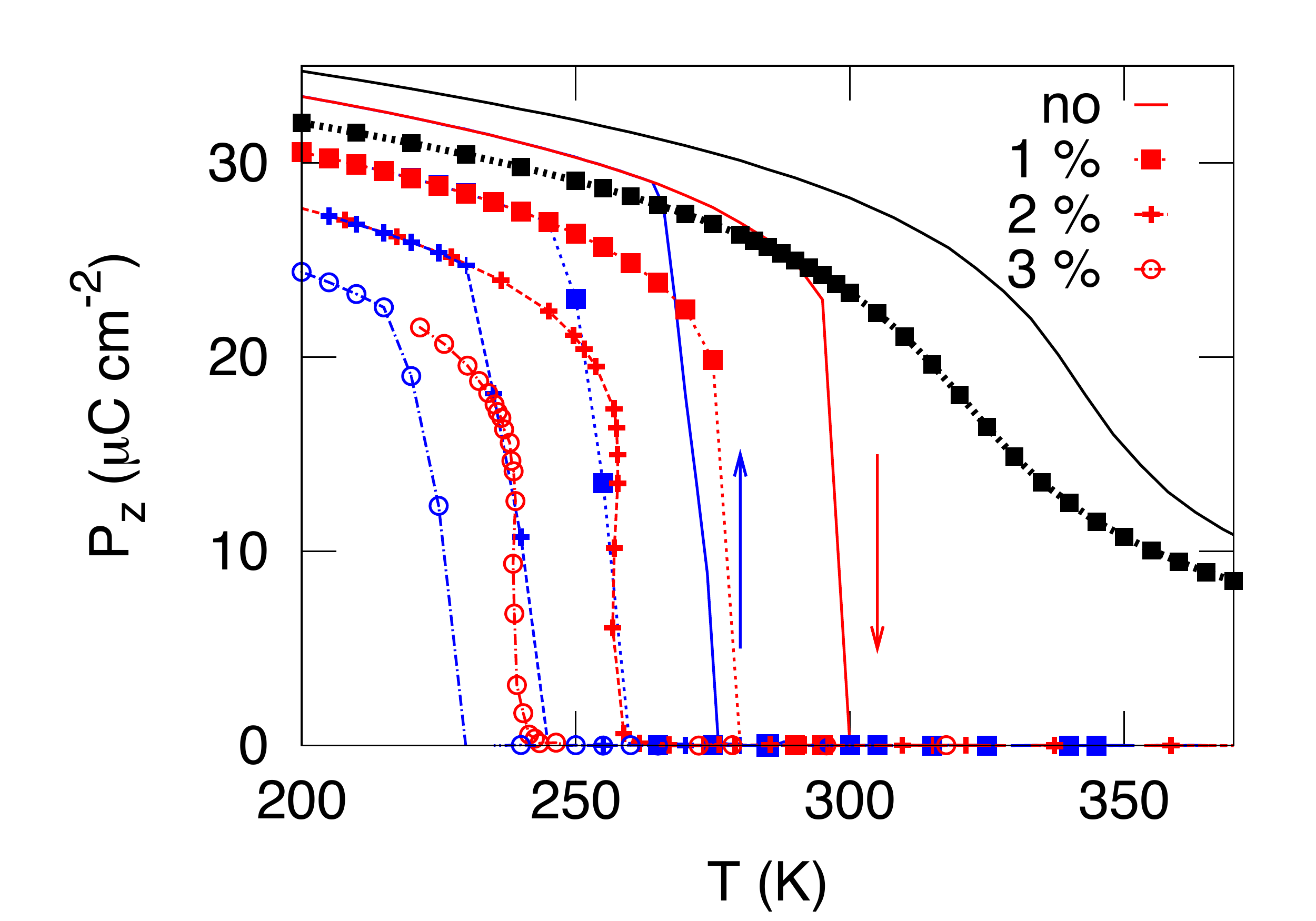}
  \caption{
    (Color online) Polarization $P_z$ of BaTiO$_3$ as function of temperature $T$ with and without non-polar defects.
    Defect concentrations ranges from 0--3\%. Upward and downward arrows indicate cooling-down (blue) and heating-up (red) simulations, respectively.
    Polarization under an external field of 100 kV/cm along z is given in black.
    As the external field is larger than the critical field strength, the field induced transition 
    does not show thermal hysteresis anymore.
  }
\label{fig:Poldefect}
\end{figure}
In addition to cooling-down and heating-up simulations  the polarization profile as obtained in 
the ECE measurement is included in Fig.~\ref{fig:Pol} in black.
 First, the material has been equilibrated in an external field of 100 kV/cm, and after the field has been removed,
 polarization and temperature have been sampled.
The polarization profile in this case mainly coincides with the heating-up simulations. 

Despite this excellent qualitative description of the material, all transition temperatures are underestimated by about 100~K within the
 model and parametrization used in our study, cf.\ Refs.~\onlinecite{Nishimatsu,Zhong}.
This may mostly by attributed to the fact that  thermal expansion is strongly underestimated in the model.
One can partly compensate for this with a semi-empirical effective negative pressure given by $p=-0.005T$~GPa.\cite{Nishimatsu} 
By this approach, $T_C$ of the paraelectric to ferroelectric transition increases to about 411/436~K under cooling-down and heating-up simulations,  thus slightly overestimating experimental values. Apart from this temperature shift, the obtained phase diagram is hardly modified. 
 Especially, the polarization profile as obtained under the ECE protocol without pressure correction (black dashed line) and the data with pressure correction shifted by 150 K (black stars) show qualitatively the same trend, see Fig.~\ref{fig:Pol}.\footnote{It has to be noted that instantaneous field removal and an explicit treatment of the acoustic degrees of freedoms have been used in connection with the pressure correction.}
All results presented in the following have been obtained without the semi-empirical correction unlike otherwise stated.
However, all qualitative trends have been cross-checked with the effective pressure correction.\\

\begin{figure}
\centering
\includegraphics[height=0.45\textwidth,clip,angle= 270,trim=.5cm 1cm 0cm 0cm]{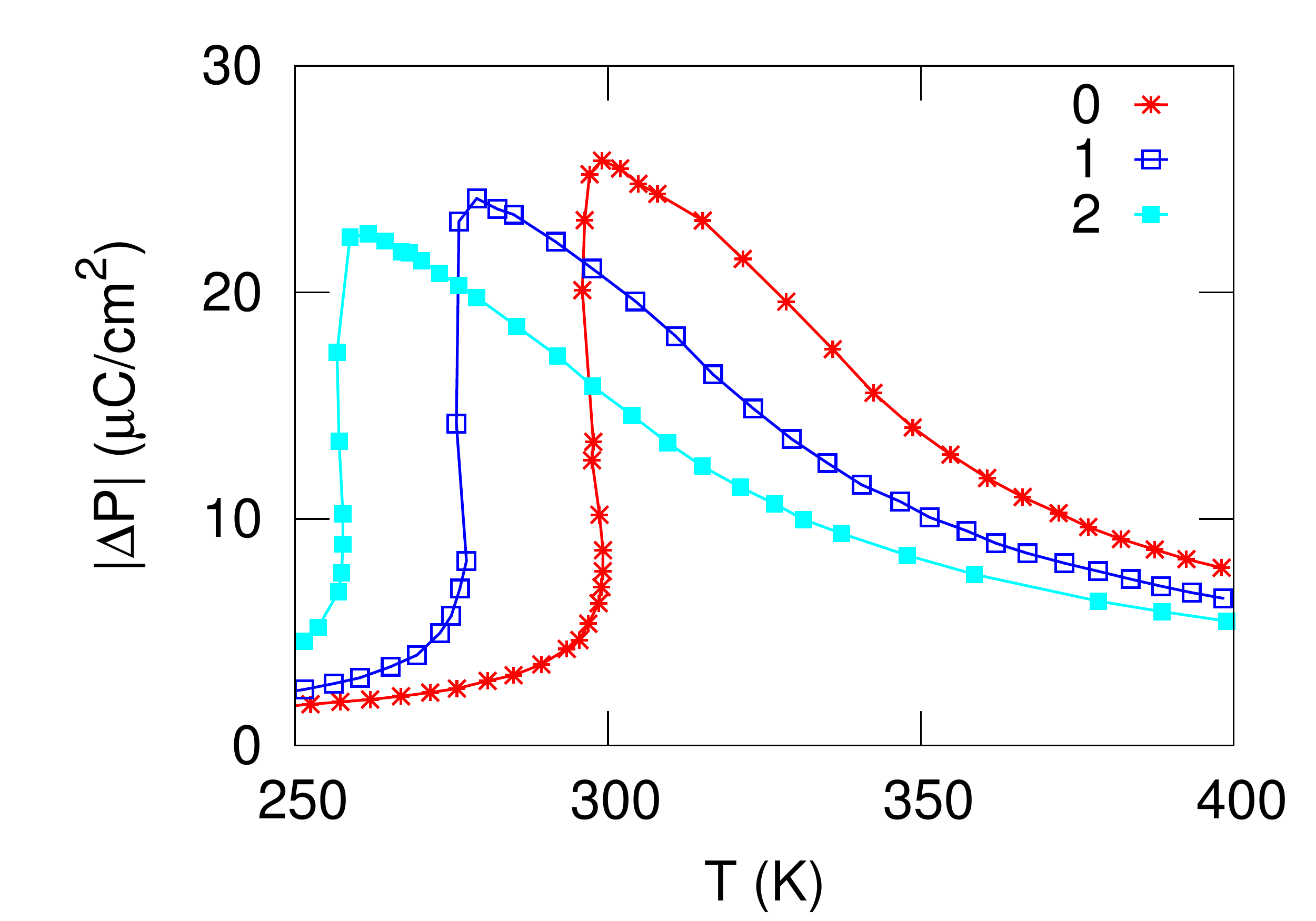}
  \caption{(Color online) Change of polarization for different concentrations of non-polar defects after an external field of 100~kV/cm
 has been  removed. 
  }
  \label{fig:ECE_defect}
\end{figure}

\begin{figure}
  \centering
  \includegraphics[height=0.45\textwidth,clip,angle=270,trim=.5cm 1cm 0cm 0cm]{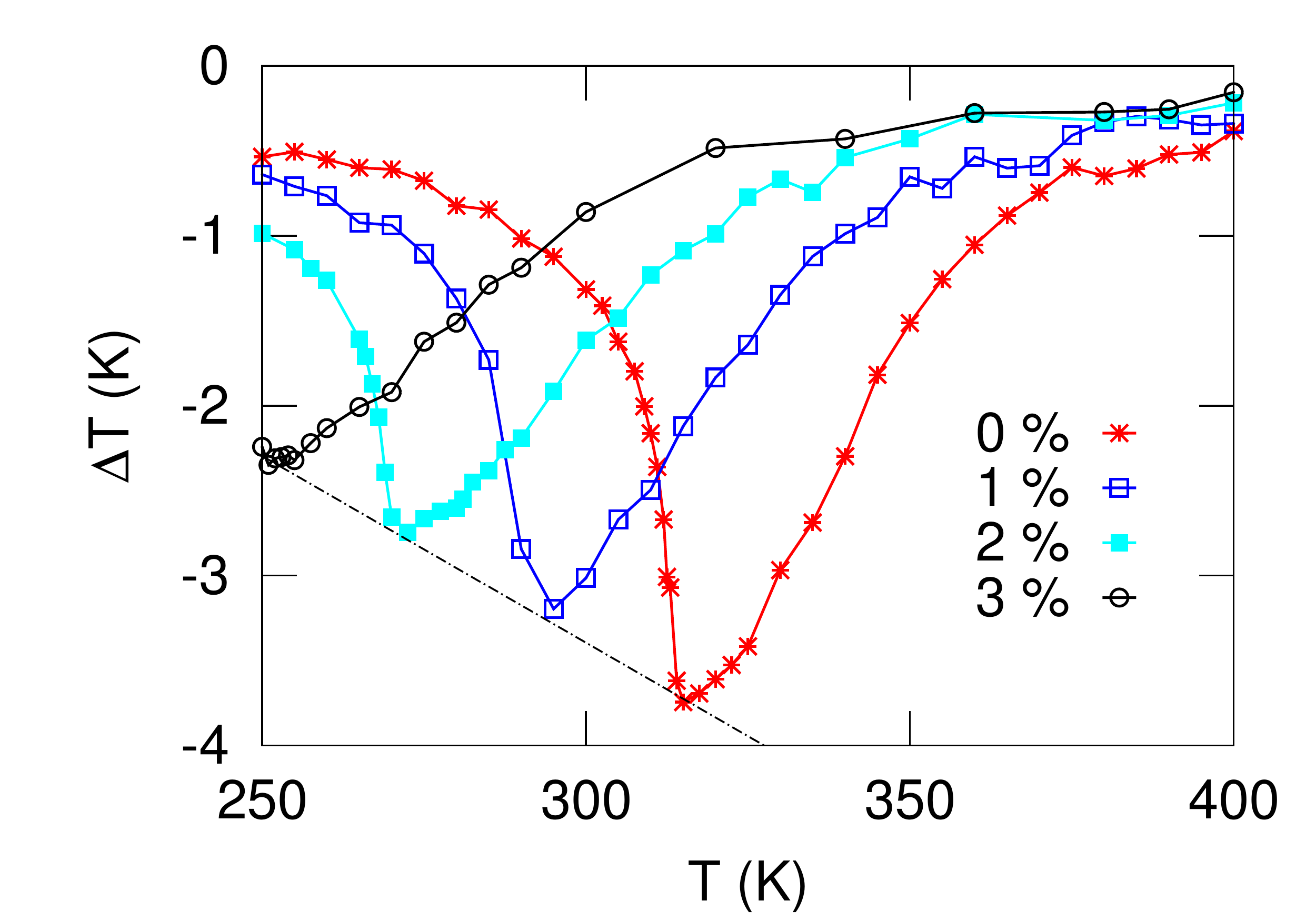}
  \caption{
    (Color online) Approximate adiabatic temperature change if an external field of 100 kV/cm is
    removed from samples with different concentration of non-polar defects. Dotted lines illustrate the linear reduction of peak position and magnitude of the ECE.
  }
\label{fig:S}
\end{figure}

The ECE in BTO can be understood by simple arguments.
Dipole ordering and magnitude of the local polar off-centering increase if an electrical field is applied, see Fig.~\ref{fig:Poldefect}. The induced
polarization vanishes, after the field is removed, see Fig.~\ref{fig:ECE_defect}. 
This reduction is related to an increase of the disorder of the local dipoles, i.e.\ an increase of the entropy under isothermal conditions, cf.\ Fig.~\ref{fig:rampkonv}~(b).
Under adiabatic conditions, this change in entropy is balanced by a
decrease of temperature, cf.\ Fig.~\ref{fig:S}. 
 The change of polarization and the corresponding adiabatic temperature
change are connected by Maxwell's relation
\begin{equation}
\frac{dT}{dE}=-\frac{T}{c_E}\left(\frac{\partial P}{\partial T}\right)_E\,.
\label{eq:dT}
\end{equation}
The polarization increases by approximately 25~$\mu$C/cm$^2$ above $T_C$ as the field of 100 kV/cm  induces a large change of polarization in this temperature range, see Fig.~\ref{fig:ECE_defect}.  In contrast, the change
of polarization by an electrical field along [001] is one order of magnitude smaller within the ferroelectric phases at
lower temperatures. 
Thus, the ECE due to the field induced ferroelectric transition above $T_C$ of the heating-up simulation is largest and a maximal adiabatic temperature change of approximately 3.7~K is obtained, see Fig.~\ref{fig:S}.
It has to be noted that the maximal adiabatic temperature change  increases to 7 K if  pressure corrections are taken into account.\cite{Nishimatsu3}  As discussed in Ref.~\onlinecite{Madhura} the ECE always increases with increasing T, cf.\ Eqn.~\eqref{eq:dT}, and thus the increase of $T_C$ by about 150~K is the most important source of this modification. 

So far, all simulations have assumed an ideal bulk material. Local reduction of polarization will occur because of defects, such as oxygen vacancies, grain boundaries and surfaces.  This setup is modeled for the limiting case of fully non-polar defects. These  "inactive" dipoles reduce the overall dipole-dipole
interaction in the system. Thus, the energy gain for a parallel alignment of all dipoles, i.e., a ferroelectric phase
with large spontaneous polarization, is systematically reduced.
As the free energy (F)
\begin{equation}
  F(T,E)=F_0(T,E)-TS(T,E)
\end{equation}
rather than the energy ($F_0$), determines the equilibrium state of the system at finite temperatures,
the transition temperature is shifted to lower temperatures, see Figs.~\ref{fig:Poldefect} and \ref{fig:ECE_defect}., since the paraelectric phase has a larger
entropy (S).
Also, the defects systematically reduce tetragonal strain (cf. Fig.~\ref{fig:strain}), strain-dipole coupling energy, and the overall
polarization  compare lines with stars and squares in Fig.~\ref{fig:ECE_defect}.
As the imposed defects may act as nucleation sites, they systematically reduce the thermal hysteresis, see Fig.~\ref{fig:Poldefect}.
The steep change of the polarization at  $T_C$ and the first order character of the transition are conserved.

As $T_C$ and $(\partial P/ \partial T)_E$ are reduced by the non-polar defects, the overall adiabatic temperature change (see Fig.~\ref{fig:S}) is systematically reduced with the number of defects, cf. Eqn.~\eqref{eq:dT}.
For example, for $\Delta E=100$ kV/cm,
5\% non-polar defects reduce the ECE by a factor of two and shift the peak of the caloric response by approximately 100~K to
lower temperatures.  The shape of the $\Delta T$(T) peak is not considerably modified by the defects. 
Since especially no broadening of the curve is visible, the temperature range with finite temperature changes is linearly reduced with the number of defects.

In summary, non-polar defects reduce the ferroelectric phase transition temperature and polarization of BTO systematically. In turn also the 
 maximal response of the ECE of BTO is reduced. However, even 5\% defects do not
suppress the caloric response and are  thus no obstacle for the use of the material in ECE applications. 
 In addition, the operation range and the temperature giving maximal adiabatic response may be
lowered by non-polar defects.
 One has to
note, that the present model overestimates the effect of local defects as  "real" defects most likely correspond to a
local reduction of the polarization rather than a complete freezing and thus a smaller shift and reduction of the
adiabatic response. 

\section{Polar complexes}
\label{sec:polar}
In the present section we discuss the influence of stable local defect dipoles. For example, this setup corresponds to dopant-oxygen vacancy complexes in transition metal doped perovskites.
It has been reported\cite{Erhart,Erhart2,Erhart3} that such complexes form immediately during the synthesis
process, that they have much slower relaxation times than the free dipoles related to the ferroelectric soft mode, and
that different relative orientations towards the spontaneous polarization of the host ferroelectric material are
possible.  In the present study, we assume fixed defect dipoles, which is a rather realistic approximation at low and
ambient temperatures during the time span of a typical  ECE measurement, cf.\ Ref.~\onlinecite{Erhart}.

Dipoles with a charging of 2 $|e|$ and distance of 1.91~{\AA} have been found for fully equilibrated O vacancy-Cu complexes in PbTiO$_3$.\cite{Erhart2} This would
correspond to a local polarization of approximately 106 $\mu$C/cm$^2$ if one defect dipole per BTO unit volume is
assumed.  Fixed local soft mode amplitudes up to 0.4~{\AA} are taken into account, corresponding to a locally enhanced 
spontaneous polarizations of 100 $\mu$C/cm$^2$ at the maximum.

\subsection{Unipolar cycling}
\label{sec:dipol}
\begin{figure}
  \centering
  \includegraphics[height=0.45\textwidth,clip,angle=270,trim=0.5cm 1.5cm 0cm 0cm]{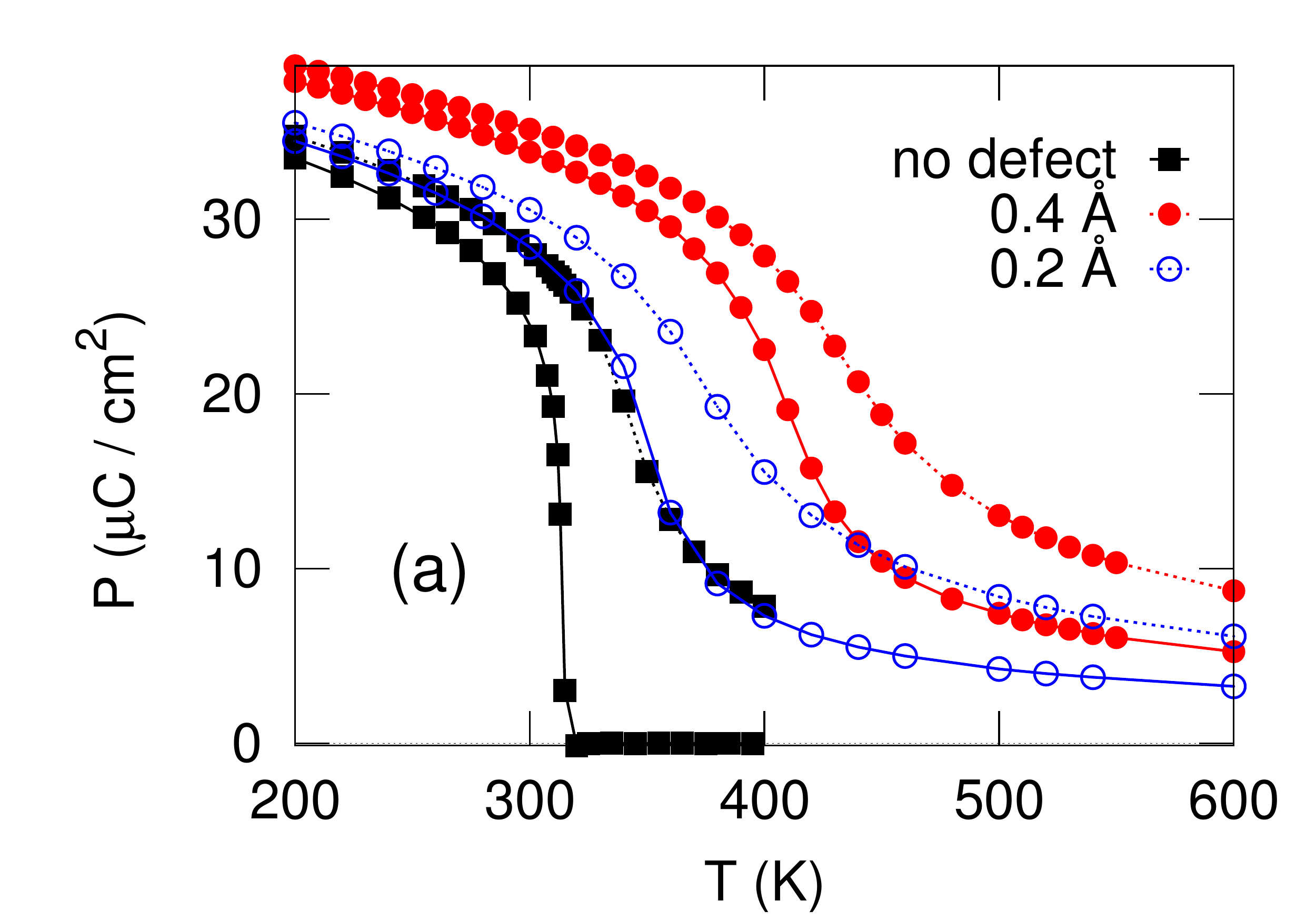}
  \includegraphics[height=0.45\textwidth,clip,angle=270,trim=0.5cm 1.5cm 0cm 0cm]{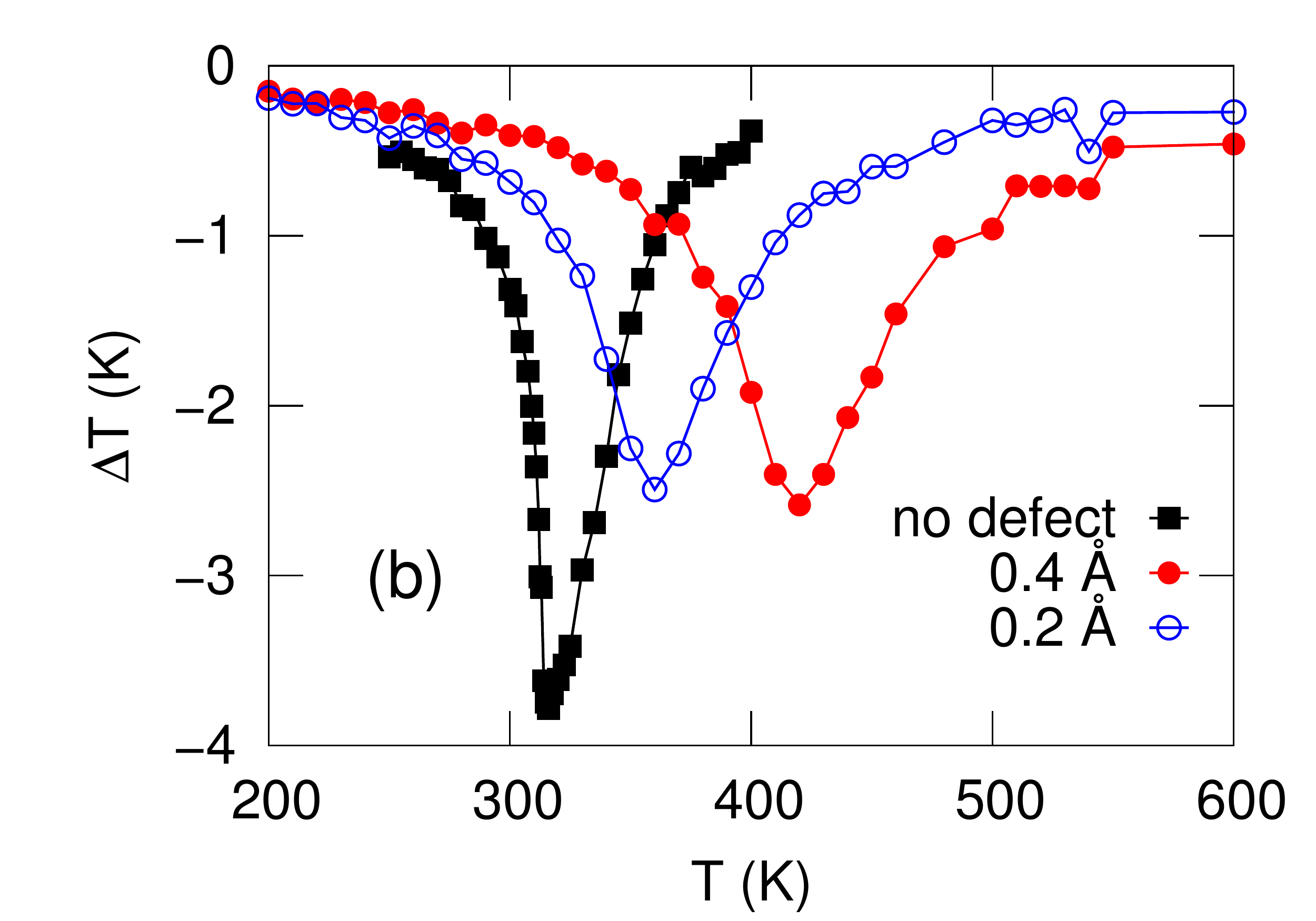}
  \caption{
    (Color online) Polarization of BaTiO$_3$ with 1\% polar defects aligned parallel to the
    external field of 100 kV/cm and after switching off the field.
    Black squares: no defects; blue open circles: 50$\mu$C/cm$^2$; red circles 100 $\mu$C/cm$^2$ (b) corresponding approximate adiabatic temperature change.
  }
  \label{fig:PTP}
\end{figure}

For ferroelectric perovskites, it has been shown that it is energetically most favorable if
the defect dipoles order with the overall polarization of the ferroelectric phase.\cite{Erhart,Erhart2} Equilibration to this ground state (aging) can be achieved, e.g., by cooling in a strong external field. If a unipolar external electrical field is used in the ECE measurement, one may assume fixed defect dipoles parallel to the direction of the field for the whole measuring time (switching on and off the field).

The influence of 1\%  defect dipoles with strengths corresponding to soft mode amplitudes of 0.2~{\AA} (50~$\mu$C/cm$^2$) and 0.4~{\AA} (100 $\mu$C/cm$^2$) on the phase diagram without and
with a parallel external field of 100 kV/cm are illustrated in Fig.~\ref{fig:PTP}.  
With increasing strength or number of defect dipoles the transition temperature is shifted to higher temperatures.
This finding is important with respect to the interpretation of experimental results on doped BTO there either an increase, no modification, or even a decrease of $T_C$ have been discussed. 
Our results show that fully equilibrated defects in a crystalline sample may increase $T_C$. In contrast to this, defects may reduce $T_c$ by extrinsic effects, such as the reduction of the grain size in experimental samples.

Already for 1\% defects with a soft mode
amplitude of 0.1~{\AA} the polarization profile is smooth and the thermal hysteresis is smaller than the accuracy of the
simulation. This soft mode amplitude corresponds to a local polarization of 25 $\mu$C/cm$^2$ which is even slightly smaller than the free dipoles of the host material.
For an amplitude of 0.2~{\AA}, we find a rather smooth change of polarization with temperature,  a vanishing thermal hysteresis, and a large polarization above the critical temperature of the ideal system. Thus, the system shows the characteristic features of a field induced
polarization above the critical field strength rather than the first order transition of the ideal system, see Fig.~\ref{fig:PTP}~(a). 
The fixed local defect dipoles create an
additional internal electrical field and thus the overall polarization and the transition temperature increase with the
strength of the defect dipoles and their density.

Similarly, an internal bias field has been found experimentally in aged doped BTO.\cite{Arlt2}
There, the hysteresis is shifted along the field axis and the coercive field increases, if the defect dipoles in a doped sample have fully ordered.

It has to be noted that the shift of the transition temperature by 100~K  and the magnitude of the internal dipole field overestimate experimental results, e.g., in Ref.~\onlinecite{Arlt2}.
On the one hand, the used model for defects neglects relaxation effects and domains. One the other hand, it is not ensured that
all dopants form fully relaxed dipole complexes in experiment, which anyway forbids a quantitative comparison of the
bias field. In addition, we have assumed a maximal defect dipole amplitude of  0.4~{\AA} corresponding to a rather large local polarization of 100~$\mu$C/cm$^2$ which may exceed the dipole moment of the real doped material.
 However, the same qualitative trends are found for a soft mode amplitude of 0.1 {\AA}. Here,
the shift of $T_C$ and thus the adiabatic response is only 10~K.
In summary, polar defects induce an internal electrical field which increases $T_C$ and the polarization while the first order character of the ferroelectric transitions weakens.

For the determination of the ECE the free dipoles are first equilibrated in an external field which is removed afterwards, see Secs.~\ref{sec:comp} and \ref{sec:nonpolar}.
During this field removal, the free dipoles relax  between the  two equilibrium phases found with and without external field as illustrated in Fig.~\ref{fig:PTP}.
Compared to the ideal system without defects, the change of polarization 
between both states is reduced by the internal field induced by the defects.
Although the system cools down under field removal, the fixed dipoles reduce the caloric
response.  In addition, the smoothing of the polarization profile reduces $\Delta T$, cf.\ Eqn.~\ref{eq:dT},  possible contributions by the latent heat of the first order transition are lost as the system is beyond its critical
point, and the number of free dipoles is slightly reduced by the defects.

The caloric response of the ideal system approximately halves if 1\% defects with a dipole moment corresponding to soft mode amplitudes between 0.1 and
0.4~{\AA} are imposed. We want to note that qualitatively the same caloric response is found by the indirect approach, i.e.\ by the integration of different equilibrium P(T,E) curves, see Sec.~\ref{sec:append}.
In the simulations, the maximal adiabatic temperature change and the shape of the $\Delta T(T)$ curve are basically the
same for different dipole strengths (0.1-0.4~{\AA}) as soon as the internal field exceeds the critical field strength.  
The reduction of the peak maximum is accompanied by a slight broadening of
the $\Delta T(T)$ curve.  
In turn a finite ECE if found for larger temperature range compared to the pure system.
As the polarization profile is shifted to higher temperatures proportional to the
strength of the internal dipole field, one could additionally adjust the optimal temperature of the caloric response by
using proper dopants.
Complementary to
the dipole strength, the density of defect dipoles modifies the induced internal field. Thus, if the number of defects
increases, the peak of the adiabatic response for an external field of 100 kV/cm is further reduced, broadened and shifted to higher temperatures.

In summary, unipolar dipole defects pointing along the direction of the external field are no obstacle for a large ECE.
Within the present model simulations an approximate ECE of 2~K can be found for an external field of 100 kV/cm for 1\% defects. In addition the
$\Delta T(T)$ curves and thus the operation range of the ECE broadens, which is beneficial for operation.  Only the
narrow peak of approximately 10 K width directly at $T_C$ is reduced as the internal bias field is beyond the critical
field.  Although the used simple model may quantitatively overestimate the shift of the caloric response with
temperature, the use of defect dipoles is clearly no obstacle and may even be used to tune the operation range of a ECE
device under unipolar cycling.

\begin{figure}
  \centering
  \includegraphics[width=0.35\textwidth,clip,angle=270, trim=.7cm 1cm .2cm .7cm]{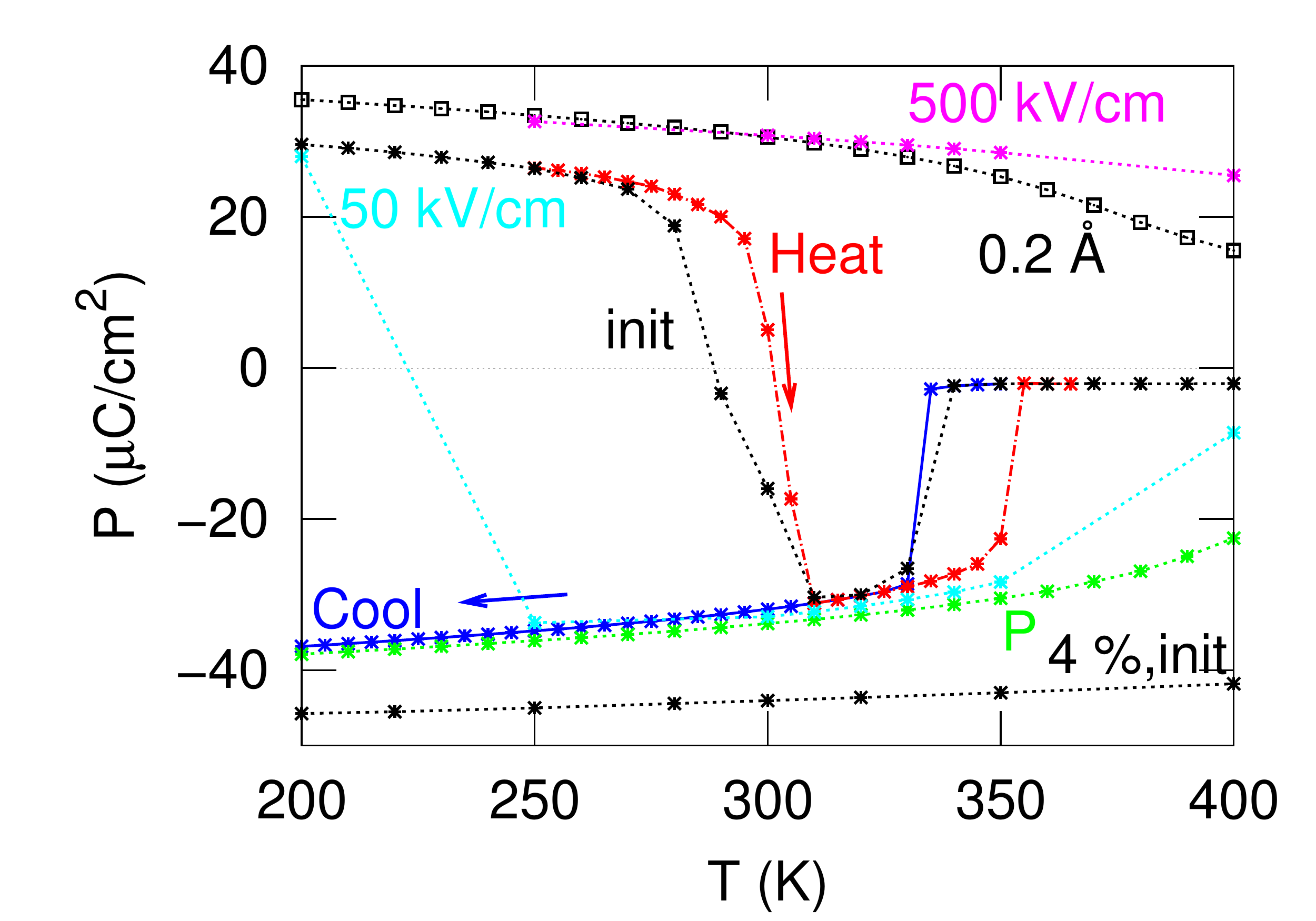}
  \caption{(Color online)  Influence of antiparallel defects on the polarization of BaTiO$_3$ for different defect densities (4\%: black curve at the bottom, 1\%: otherwise), strengths (local soft mode amplitude of 0.4~{\AA}/0.2~{\AA}: stars/squares)
     and different external fields (500~kV/cm: magenta; 50 kV/cm: cyan; 100 kV/cm otherwise) for cooling-down simulations.
     For 1\% defects with a strength of 0.4~{\AA} also results for heating-up (red) and initalized from scratch (black) as well as cooling-down simulation for parallel defects (green, with label $P$) are given. 
  }
  \label{fig:P_down_field}
\end{figure}
\subsection{Antiparallel defects}
\begin{figure}
  \centering
 \includegraphics[width=0.5\textwidth,clip, trim=.5cm .7cm 1.2cm 0.7cm]{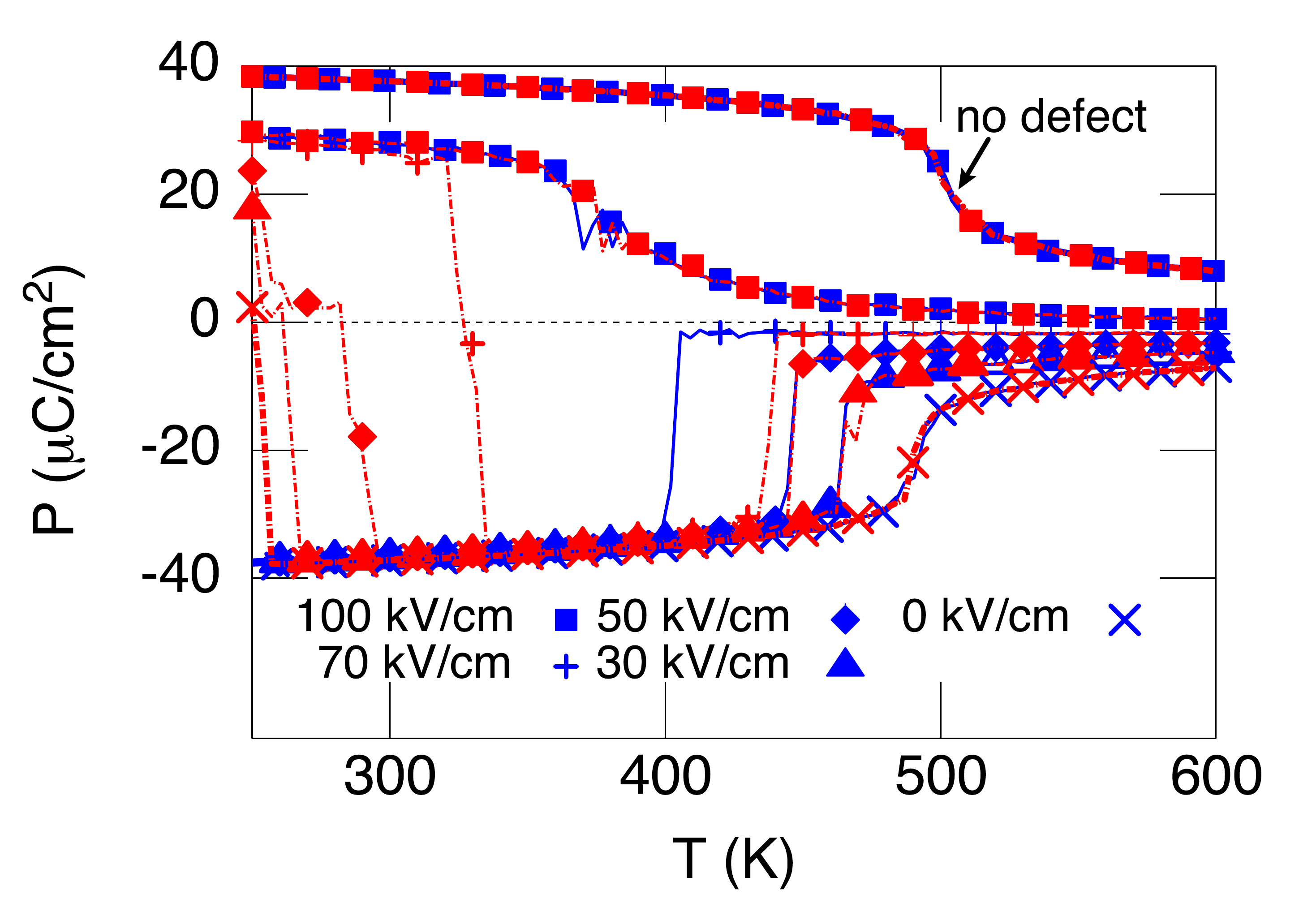}
  \caption{
    (Color online) 
 Polarization of BTO for 1\% defects with a soft mode amplitude of 0.2~{\AA} under cooling down (solid blue lines) and heating up (dashed red lines)  for different field strength. In contrast to Fig.~\ref{fig:P_down_field} a temperature dependent expansive pressure has been applied. A 16x16x16 supercell only has been used.
  }
  \label{fig:P_down_fieldb}
\end{figure}

\begin{figure}
  \centering
  \includegraphics[height=0.5\textwidth,clip,angle=270,trim=.5cm 1.5cm 0cm 0cm]{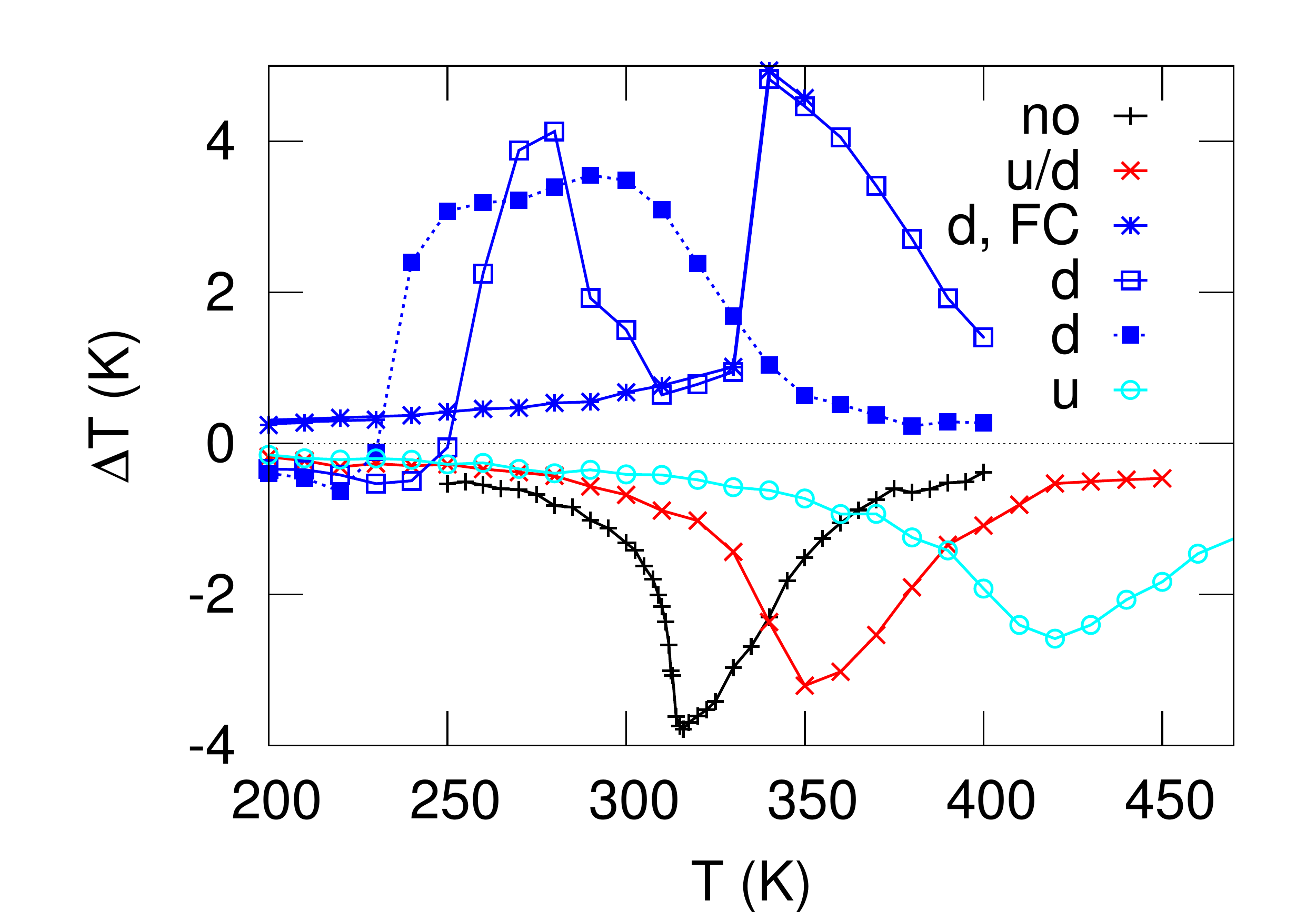}
  \caption{
    (Color online) Approximate caloric response of BTO with 1\% dipole defects if an external field of 100 kV/cm is removed.
    Black: no defects; cyan: parallel defects (u); Red: 50:50 parallel and antiparallel defects (u/d); blue: antiparallel defects (d) with soft mode amplitudes of 0.4~{\AA} (solid lines) and 0.2~{\AA} (dashed lines);
    For the 0.4~{\AA} amplitude also the caloric response after field cooling (FC) is shown.
  }
  \label{fig:P_T_AP}
\end{figure}

Depending on measurement protocol and history of the sample,
one can also think of metastable defect dipoles pointing antiparallel to the overall polarization for a certain number
of unipolar field cycles.\cite{Erhart,Erhart2} Such a setup may have exciting consequences for the phase diagram, see Figs.~\ref{fig:P_down_field} and \ref{fig:P_down_fieldb}, and the
caloric response of the material, see Fig.~\ref{fig:P_T_AP}.

We start our discussion with the influence of competing internal ($E_{\text{int}}$) and external ($E_{\text{ext}}$) fields on the phase diagram of BTO.
In the limit of an external field strongly exceeding the internal one, the free dipoles order along the external field direction (state 1), see Figs.~\ref{fig:P_down_field},~\ref{fig:P_down_fieldb}  and Tab.~\ref{tab:daten}.
In this field range, the internal field acts as a small disturbance of the ferroelectric state, only.
Small chains of free dipoles several lattice constants below and above the defect dipoles are coupled to the internal field resulting in a
slight reduction of $P$ and the tetragonal strain compared to the defect free material within an external field. 
With decreasing ratio between external and internal field, polarization and transition temperature of the ferroelectric state 1 are systematically reduced, see Fig.~\ref{fig:P_down_fieldb}.

In the opposite limit without external field, the internal field stabilizes the ferroelectric phase with polarization parallel to the defects (state 2), as discussed in the last section. 
With an increasing strength of the external field, this  state (2) is destabilized, i.e.\ the polarization of the ferroelectric phase and its transition temperature systematically reduced, see Figs.~\ref{fig:P_down_field} and \ref{fig:P_down_fieldb}.
For comparable internal and external field strengths, the stability of the ferroelectric states 1 and 2 depends also on the history of the sample and the simulation protocol, see the example of  a soft mode amplitude of 0.2~{\AA} in  Fig.~\ref{fig:P_down_fieldb}.
First of all, the competing fields open a thermal hysteresis between cooling-down and heating-up simulations for state $2$.
Although, state $2$ has the lowest energy and is found in  cooling-down simulations up to an external field strength of  70 kV/cm, state $1$ is at least a  metastable state for $E_{\text{ext}}\geq$ 30~kV/cm and is found for heating-up simulations or after field-removal.

Below the transition temperature T$_C$, internal and external field occasionally compensate each other, resulting in a state without overall polarization (state 3).
For example, for a soft mode amplitude of 0.4~{\AA} and an external field of 100~kV/cm a polarization of  approximately 2~$\mu$C/cm$^2$ is observed above 325~K/350~K (for cooling-down/ heating-up), which  is
twice the polarization of  pure frozen defect dipoles, see Fig.~\ref{fig:P_down_field}. A snapshot of the local dipole configuration at 400~K
reveals that the distribution of local soft mode amplitudes shows two maxima at approximately 17 and -13~$\mu$C/cm$^2$.  Parts of the free dipoles align with external, while a similar number of dipoles near the defects align with internal defect field. 

It has to be noted that all three states, their different temperature dependency and their dependency on the sample history can be reproduced with different
random defect distributions, cell sizes, and defect strengths, see Tab.~\ref{tab:daten}. In addition, the use of an external pressure does not modify the phase diagram despite the increase of transition temperature and polarization which as also been discussed for ideal BTO, cf.~Figs.~\ref{fig:P_down_field},\ref{fig:P_down_fieldb}, and Tab.~\ref{tab:daten}.

In summary,  the stability region of three phases with pronounced differences in dipole ordering and polarization, can be adjusted by the relative strength of the external field and the internal field given by density and magnitude of the defect dipoles. In addition, different metastable states can be found depending on the simulation protocol and the history of the sample for similar strength of internal and external field.\\

\begin{table}
\caption{Exemplary phase sequences found for different relative strength of internal ($E_{\text{int}}$ as given by dipole strength and concentration) and external fields ($E_{\text{ext}}$) found for defect dipoles antiparallel to the external field in cooling-down and heating-up simulations. \label{tab:daten}}
\begin{tabular}{cccc}
\hline
\hline
&field & phase sequence  \\
\multirow{2}*{(a)}&\multirow{2}*{E$_{\text{ext}} >E_{\text{int}}$}& cool: $P=0 \Rightarrow P || E_{\text{ext}}$  \\
&&heat: $P || E_{\text{ext}}  \Rightarrow P=0 $ \\
\multicolumn{4}{l}{e.g., 1 \% 0.4 {\AA} 500 kV/cm, 1\% 0.2 100 kV/cm}\\
\hline
\multirow{2}*{(b)}&\multirow{2}*{E$_{\text{ext}} \sim E_{\text{int}}$}& cool: $P=0\Rightarrow P||E_{\text{int}}$ \\
&&heat: $P||E_{\text{ext}} \Rightarrow P || E_{\text{int}} \Rightarrow P=0$\\
\multicolumn{4}{l}{e.g., 1\% 0.4 {\AA} 100 kV/cm, 1\% 0.2 {\AA} 70 kV/cm}\\
\hline
\multirow{2}*{(c)}&\multirow{2}*{E$_{\text{ext}} <E_{\text{int}}$ }& cool: $P=0\Rightarrow P|| E_{\text{int}}$ \\ 
&&heat: $P||E_{\text{int}}\Rightarrow P=0$\\
\multicolumn{4}{l}{e.g., 1\% 0.4 {\AA} 50 kV/cm, 4\% 0.4 {\AA} 100 kV/cm},\\
\multicolumn{4}{l}{1\% 0.2 {\AA} 30 kV/cm}\\
\hline
\hline
\end{tabular}
\end{table}

The rich phase diagram found for anti-parallel defects has exciting consequences on the ECE which will be discussed in the following.
First of all, we want to highlight the dependency of the obtained phases on the history of the sample as discussed in the last paragraph. 
Because of this, also the caloric response depends crucial on the sample history, the equilibration and the direction of the field change.
As a consequence of this, the indirect method, i.e.\ the integration of one general set of equilibrium polarization curves cannot be used similarly to the findings for relaxors.\cite{Lu2}

We start our discussion with the caloric response under field-removal. First the system is equilibrated at each temperature within an external field. Approximately, the obtained phase corresponds to the heating-up phase diagram, see Fig.~\ref{fig:P_down_field}.
If the external field is removed, it is energetically most favorable for the free dipoles to align parallel to the defects, i.e.\ if the state discussed in the last section for parallel dipoles is reached.
However, with decreasing temperature the relaxation slows down and thus the system with polarization along the field
direction, may be stuck in this metastable state for typical simulation times. For example, after 500~ps the system
relaxes to the energetic ground state at 210~K for defects with 0.4~{\AA} amplitude after the external field of 100~kV/cm has been removed.

For the fully relaxed system, field removal changes the polarization
considerably. Especially, for initial states $1$ and $3$, a large change of  $P$ under field removal induces a large temperature change, see Fig.~\ref{fig:P_T_AP}. For state 2, with a low  coupling with $E_{\text{ext}}$,  the change of $P$ and thus the adiabatic temperature change
is reduced albeit a finite  caloric response also appears in this temperature region. 

In contrast to the defect configurations discussed in Secs.~\ref{sec:nonpolar} and \ref{sec:dipol}, the
polarization is larger without field.  Thus,  an inverse ECE is found as the system heats up under field
removal.  The effect is rather large and especially for the initial state $3$ the magnitude of the caloric response is exceeding the conventional ECE found for pure BTO at the same strength of the external field, see Fig.~\ref{fig:P_T_AP}.
It should be noted that the peak of the ECE around 350~K,  related to state $3$, does neither depend on the ramping rate, see discussion in Sec.~\ref{sec:append}, nor on the history of the sample and is found for field-on and field-off simulations, see Fig.~\ref{fig:rampkonv}~(b), as the compensation between external and internal field is found uniquely between cooling-down and heating up simulations, cf.~Figs.~\ref{fig:P_down_field} and ~\ref{fig:P_down_fieldb}.

Note, that for a dipole strength of 0.4~{\AA} also an inverse ECE of similar magnitude is found below 300 K, if an external field of 100~kV/cm is removed from the initial state $1$.
 Similar $\Delta T$ values can even be found below 250~K for extended relaxation time (not shown in Fig.~\ref{fig:P_T_AP}.)  However, for this combination of internal and external field, the metastable state 1 only appears for field-heated or un-poled samples and thus the ECE is not reversible and can only be found under field removal, cf.~Fig.~\ref{fig:rampkonv}~(b).
  After the system has reached the stable state 2 with
polarization parallel to the defect dipoles in one of the first field cycles the caloric response is reduced below 1~K in the following ones, see blue stars in Fig.~\ref{fig:P_T_AP}. 
The temperature range with a large  inverse ECE might be extended if the internal field strength is reduced, e.g., if a soft mode amplitude of 0.2~{\AA} is assumed, see Fig.~\ref{fig:P_T_AP}.
 In this case, states $3$ and $1$ are stable within the external field of 100 kV/cm down to approximately 160~K, i.e.\ down to the second ferroelectric transition temperature between the tetragonal and orthorhombic ferroelectric phases. Uniquely, the same states are reached for cooling-down and heating-up simulations and thus a reversible effect has to be expected.
 After field removal, the free dipoles align with the internal field induced by the defects if a sufficient relaxation time is taken into account, cf.\ discussion for the soft mode amplitude of 0.4~{\AA}. As the polarization in the initial states is smaller than in the final state, an inverse ECE with similar magnitude as the conventional ECE of the ideal system is found in a
broad temperature range below 350 K, see Fig.~\ref{fig:P_T_AP}. Even below 230~K a comparable caloric response has to be expected for an extended equilibration time after field removal.
\begin{figure}
  \centering
  \includegraphics[height=0.35\textwidth,clip,trim=3cm 2cm 1.5cm 2.5cm]{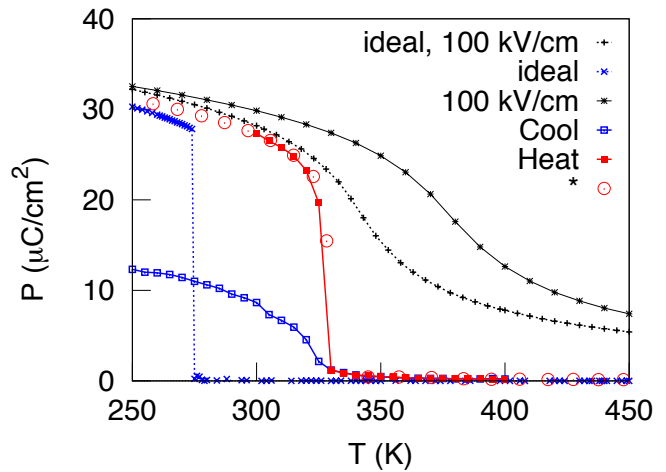}
  \caption{
    (Color online) Polarization for 1\% polar defects with equal up/down
    distribution under different simulation protocols (cooling: blue, heating and after field removal: red) and within 100 kV/cm (black). For comparison P(T) under cooling is given for the ideal system (dotted lines).\\
*: After field removal.
  }
  \label{fig:bipolar}
\end{figure}

In summary, we find an inverse ECE  in doped BTO  if the external field and
the internal field given by the defect dipoles are antiparallel.
The relative strength between
external field and internal field can be used in order to stabilize one or the other phase within the external field,
thus adjusting the magnitude and even the sign of the ECE.
Especially for similar magnitudes of external and internal field (state 3) a reversible inverse ECE is found over a broad temperature range.
 The magnitude of this  temperature change even 
 exceeds the conventional ECE we find without defects.

We note that the appearance of the large inverse ECE in doped ferroelectrics has been confirmed by simulations employing a Landau-type potential during the referee process of this paper.\cite{Ma}

\subsection{Bipolar cycling}
Most device concepts so far are based on the bipolar cycling of an electrical field.
For such a setup ``de-aging'' of the sample is present, i.e.\ the internal bias
field discussed for fully polarized samples with defects, vanishes with time.  It has been discussed in literature that the slow
defect dipoles cannot follow the electrical field and are thus exposed to a surrounding polarization pointing half times up and down.
 As consequence, also half the defect dipoles point in either field direction.\cite{Erhart}

In the following, we assume fixed defect dipoles pointing randomly either parallel or antiparallel to the external field
with a 50:50 distribution.  For 1\% defects corresponding to a local soft mode of 0.4~{\AA} the transition temperature
under cooling-down increases by approximately 30~K compared to the ideal system, see Fig.~\ref{fig:bipolar}.
\begin{figure}
  \centering
  \includegraphics[height=0.35\textwidth,clip,trim=1cm 0.2cm 0cm 0.7cm]{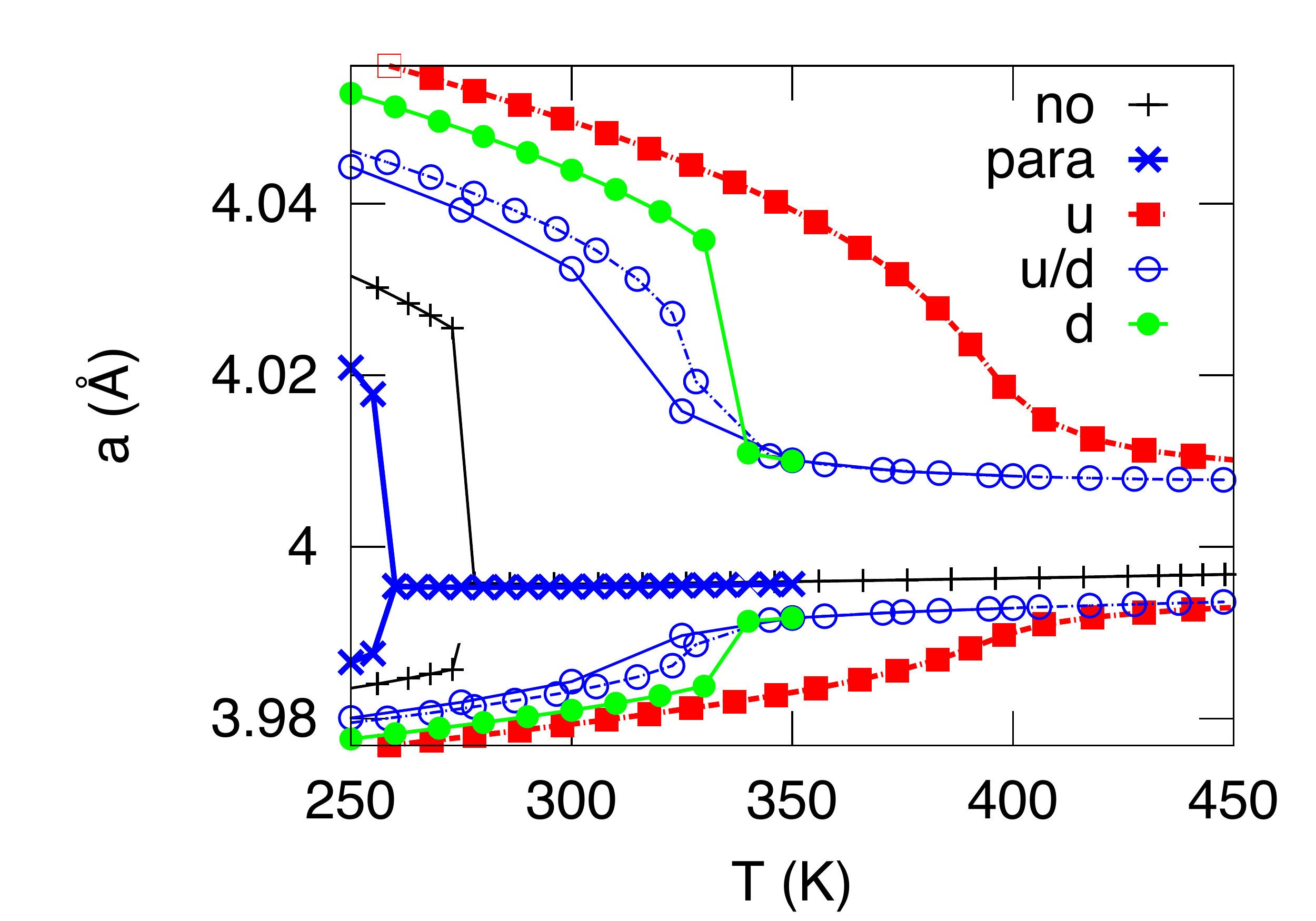}
  \caption{(Color online) Illustration of the temperature dependent lattice constants along and perpendicular to the tetragonal axis for the ideal system (black), 
    1\% non-polar defects (blue crosses), and 1\% polar defects with 0.4 {\AA} amplitude (dashed lines).
    Empty squares: no external field; Empty circles: 50:50 distribution of parallel and antiparallel dipoles;
    Filled circles: dipoles antiparallel to an external field of 100~kV/cm. Solid lines: cooling-down and dashed lines: after field removal.
  }
  \label{fig:strain}
\end{figure}

Furthermore, the local defects act as kind of precursor and thus the thermal hysteresis between cooling-down and
heating-up simulations is reduced to at most 5~K.  It has to be noted that the overall polarization is reduced under
cooling for the configuration shown in  Fig.~\ref{fig:bipolar}, as 180$^{\circ}$ domains with up and downwards
polarization are formed.  However, the local soft mode, as well as the tetragonal ratio is very similar between
cooling-down and heating-up, see Fig.~\ref{fig:strain}.

One important mechanism for the shift of $T_C$ can be found in the coupling between
local strain and local soft mode.  Thus, a tetragonal distortion is induced near the polar defects i.e.\ the lattice
constant along the local dipole direction increases compared to the ideal system for all temperatures, see
Fig.~\ref{fig:strain}.  This tetragonal distortion stabilizes the ferroelectric phase, the energy barrier for
the ferroelectric transition is reduced, and $T_C$ as well as the saturation polarization increase.
If the in-plane lattice constants are clamped to their bulk value, the defect induced strain is reduced and also the shift of $T_C$ to higher temperatures does not show up anymore in simulations of the material with frozen defects.

As no net internal field builds up for the equal distribution of both kind of defects, the polarization shows a steep
jump at $T_C$ for heating-up simulations or the ECE protocol.  Analog to the ideal system, an external field induces a
spontaneous polarization, resulting in large adiabatic temperature change above the transition temperature.
However, the dipoles slightly reduce the adiabatic response at its peak maximum. This effect is similar to 1\%
non-polar dipoles but a slightly broader peak is found. 

In summary, for all relative orientations between local defects and external field, taken into account in this investigation, 
we find caloric responses which are comparable to the response of ideal BTO.
  Especially, the
dipole distribution under an cycling external field is no obstacle for the ECE. The operation range is even broadened
in comparison to the ideal system and it is slightly shifted to higher temperatures.  

\begin{table*}
\caption{Summary of the influence of different kind of defects on the phase diagram and caloric response as obtained within our {\it{ab initio}} based molecular dynamics simulations. \label{tab:sum}}
\begin{tabular}{ccccc}
\hline
Impurity & T$_C$ & ECE&thermal hysteresis\\
non-polar defect or & \multirow{2}{*}{reduced}& reduced &\multirow{2}{*}{reduced}\\
 locally reduced polarization &&no broadening\\
\hline
polar defects  &  \multirow{2}{*}{increase}&reduced&\multirow{2}{*}{vanishes} \\
parallel to external field &&broadening of $\Delta T(T)$\\
\hline
50:50 distribution of  &  \multirow{2}{*}{increase} &  reduced& \multirow{2}{*}{vanishes}\\
parallel and antiparallel defects && broadening of $\Delta T(T)$\\
\hline
antiparallel defects & \multirow{2}{*}{increase}  &inverse&different metastable\\
in unipolar field&&broad range&states \\
\hline
randomly ordered & \multirow{2}{*}{reduced}&\multirow{2}{*}{reduced}&\multirow{2}{*}{reduced}\\
dipoles\\
\hline
\end{tabular}
\end{table*}
\begin{figure}
  \centering
  \includegraphics[height=0.35\textwidth,clip,trim=0cm .6cm 0cm .7cm]{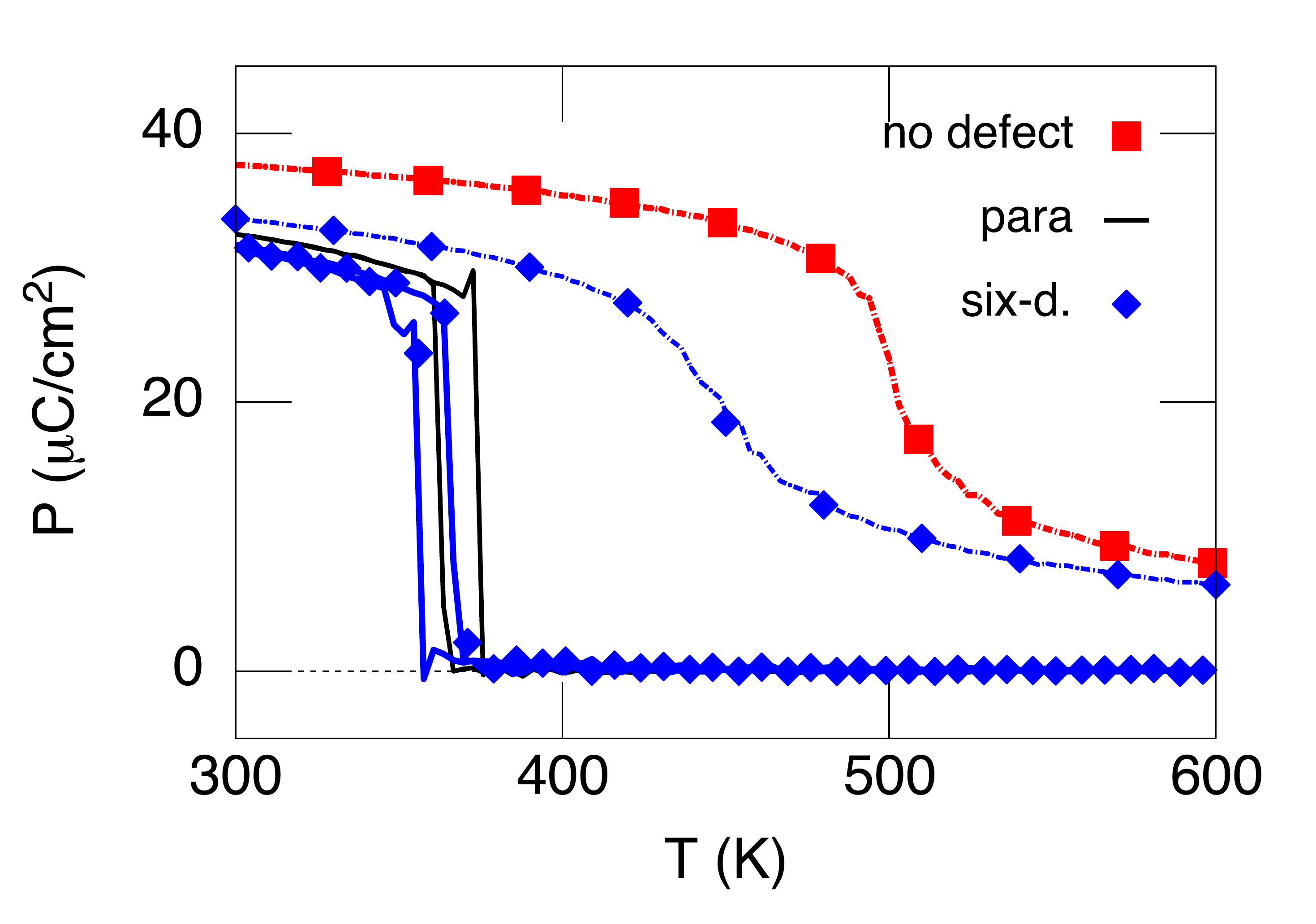}
  \caption{
    (Color online) Cooling-down and heating-up simulation including a temperature dependent pressure correction for a 16x16x16 simulation cell.
 1\% defects corresponding to a soft mode amplitude of 0.2~{\AA} have been taken into account for dipoles pointing randomly along the six cubic crystal directions (blue diamonds) and and non-polar defects (black). The polarization profiles without external field (solid lines) and in an external field of 100~kV/cm along z (dashed lines) are opposed.
  }
  \label{fig:ECE_random}
\end{figure}
\subsection{Disordered dipoles}
\label{sec:random}
For the as-prepared  cubic sample, the alignment of defect dipoles along all cubic lattice direction has the same energy and
probability. 
In the following we discuss the influence on the phase diagram and the ECE by defect dipoles perpendicular to the applied field.

The influence of a uniform distribution of 1\% dipoles corresponding to  a soft mode amplitude of 0.2 {\AA}  along all cubic directions on the phase diagram is illustrated in Fig.~\ref{fig:ECE_random}.
Heating-up and cooling-down simulations including a temperature dependent pressure correction have been performed. The jump of the polarization at $T_C$ is steep and a thermal hysteresis is present, i.e.\ the first order character of the FEL transition is conserved. Here, only one sixth of the dipoles point parallel or antiparallel to the external field. This results in rather small internal
fields along a certain direction. In addition, the dipole fields should in average cancel each other.  
In contrast to the example of dipoles pointing along and antiparallel to the overall polarization, no direction for the
tetragonal strain is favored by the defect dipoles and thus $T_C$ is not shifted to higher temperatures.
Instead, the magnitudes of $\partial P/\partial T_{E}$ and $T_C$ are reduced by the doping similar to 1\% non-polar defects and thus a slightly
reduced ECE has to be expected.

In summary, the equal distribution of defect dipoles along all cubic lattice directions in an as-prepared cubic sample may slightly reduce the ECE. However after several ECE cycles, the defect dipoles would most likely equilibrate with the direction of the applied field\cite{Erhart,Erhart2,Erhart3} and the ECE as discussed in Sec.~\ref{sec:polar} would recover.

\section{Conclusions and Outlook}
\label{sec:sum}
We have discussed the influence of different defects on the ferroelectric and electrocaloric properties of BTO in the framework of {\it{ab initio}} based molecular dynamics simulations.  The main influences of different kinds of defects are summarized in Tab.~\ref{tab:sum}.
We find that defects with locally reduced polarization tend to reduce $T_C$ and the adiabatic response. The same trend is found for strong defect dipoles with slow relaxation, but without a preferred alignment.
However, in both cases the reduction of the ECE is only small for the tested defect concentrations and  thus does not contradict a use of BTO in ECE cooling devices.

In addition, we have discussed the possible benefit of dopants for the ECE. For this purpose we have simulated the influence of fixed defect dipoles on the ferroelectric phase diagram and the ECE systematically.
Dipoles pointing along the tetragonal axis of BTO tend to increase $T_C$ which could partly be related to the strong dipole-strain coupling. In addition, parallel dipoles induce an internal electrical field which results in an increased polarization and reduces the first order character of the ferroelectric transition.

If the defect dipoles, or at last half of them, are pointing parallel to the external field, the caloric response is slightly reduced and the $\Delta T$ curve is shifted to higher temperatures and broadens. This kind of defect dipoles may be beneficial for applications based on ferroelectric materials as they open up a possibility to broaden the operation range.
Excitingly, the doping of BTO samples may even induce different stable and metastable ferroelectric states and an inverse ECE under certain measurement protocols. Our investigation has shown that defect dipoles pointing antiparallel to the external field form an internal electrical field which couples to the free dipoles with a different temperature dependency than the external field.
The relative strength between internal and external field at a certain temperature may be adjusted, e.g., by the number and strength of the imposed defect dipoles. 
As a consequence three different ferroelectric states with polarization along external field,  internal field, or nearly compensated polarization can be seen in our simulations.
 
 If the field is removed, the free dipoles align again with the defect dipoles resulting in an increasing polarization and thus a heating of the sample for all these states.
 Especially the high-temperature state which we found for different combinations of internal and external field is promising for the modification of the ECE:
In our simulations the polarization within the external field is quenched in a broad temperature interval below $T_C$, resulting in a large change of the polarization and a large caloric response under field removal.

  Future work will concentrate on improved modeling of the influence of defects, e.g.\ by using slowly relaxing but dynamic defect dipoles and by taking the coupling between external field and defect dipole strength into account. In addition, the combination of defects with alloying is promising with respect to adjusting the operation temperature for possible applications.

\section*{Acknowledgements}
Financial support has been guaranteed by the Deutsche
Forschungsgemeinschaft via the SPP 1599.
The work of TN was supported in part by JSPS KAKENHI Grant Number 25400314 and 26400327.
This work was also supported in part by the
Strategic Programs for Innovative Research (SPIRE), MEXT,
and the Computational Materials Science Initiative (CMSI), Japan.
Computational resources were provided by the Center for Computational Science and Simulation (CCSS), University of Duisburg Essen and  
 the Center for Computational Materials
Science, Institute for Materials Research (CCMS-IMR), Tohoku University.
We thank the staff at CCMS-IMR for their constant effort.
This research was also conducted using the Fujitsu PRIMEHPC FX10 System (Oakleaf-FX, Oakbridge-FX)
in the Information Technology Center, The University of Tokyo.
We want to acknowledge the fruitful discussion with Claude Ederer, Vladimir Shvartsman, Karsten Albe, Doro Lupascu, Marianne Schr\"oter, and Peter Entel.

\section{Appendix}
\label{sec:append}
In the following, technical aspects of the simulation will be discussed in detail, a comparison of the direct and indirect method will be reviewed from literature,\cite{Cp} and additional convergency tests will be presented.
In the present paper we calculate the caloric response of BTO directly, see Sec.~\ref{sec:comp}. For this purpose we equilibrate the system at a given temperature ($T_1$) and external field ${\cal{E}}_1$, ramp the field to its final value ${\cal{E}}_2$, and monitor the change of the kinetic energy ($T_2)$.
By this approach we can get a good qualitative description of the change of the ECE with different kind of defects as we take the most relevant degrees of freedom into account and induce similar errors without all our simulations.\cite{Cp,Ponomareva,Nishimatsu3,Lisenkov}

Alternatively, the thermodynamic Maxwell relations (cf.\ Eqn.~\eqref{eq:dT}) can be used to calculate the caloric response by integration of the pyroelectric response at different values of the external field (indirect method) 
\begin{equation}
\Delta T=-\int_{{\cal{E}}_1}^{{\cal{E}}_2}\frac{T}{C_{p,\cal{E}}}\left(\left.\frac{\partial P}{\partial T}\right)\right|_{{\cal E}} d {\cal E}\,.
\end{equation}
Here, $C_{p,{\cal E}}$ is the specific heat at constant pressure and field, and the external field is varied form ${\cal E}_1$ to ${\cal E}_2$.
This indirect approach breaks down at the first order phase transition as $\partial P/\partial T$ is ill-defined, the specific heat diverges, and possible contributions stemming from the latent heat are not accounted for.\cite{Cp}
In addition, the indirect method fails if the caloric response depends on the simulation protocol, the history of the sample, or the direction of the field change (on-off), e.g., in relaxors.\cite{Lu2}
However, the indirect approach has the important advantage that only equilibrium states are taken into account and it can be used in order to cross check the results obtained via the direct approach, which include two important approximations.

\begin{figure}
\includegraphics[width=0.5\textwidth,clip,trim=2cm 2cm 0cm 2cm]{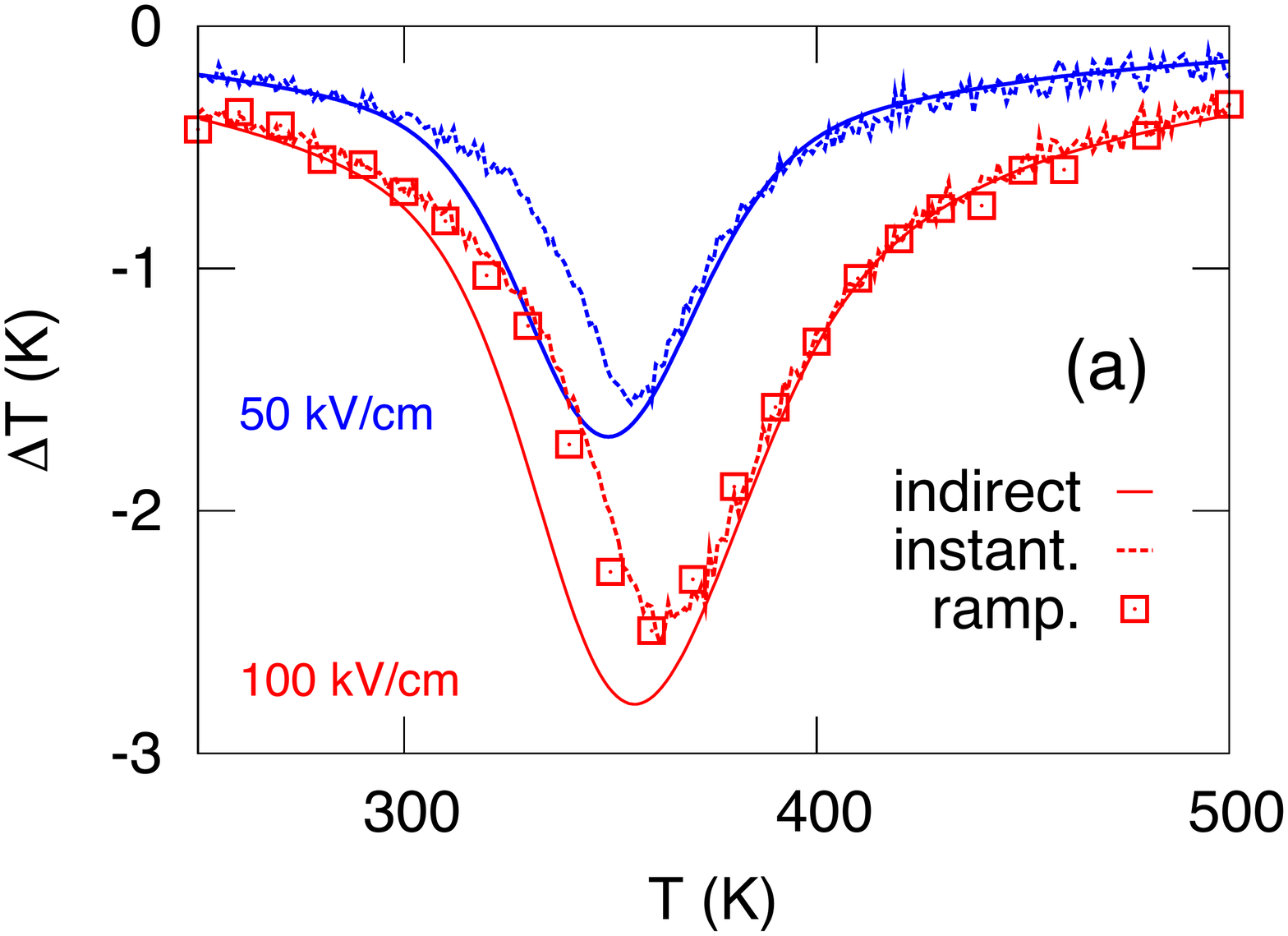}
\includegraphics[width=0.5\textwidth,clip,trim=2cm 2cm 0cm 2cm]{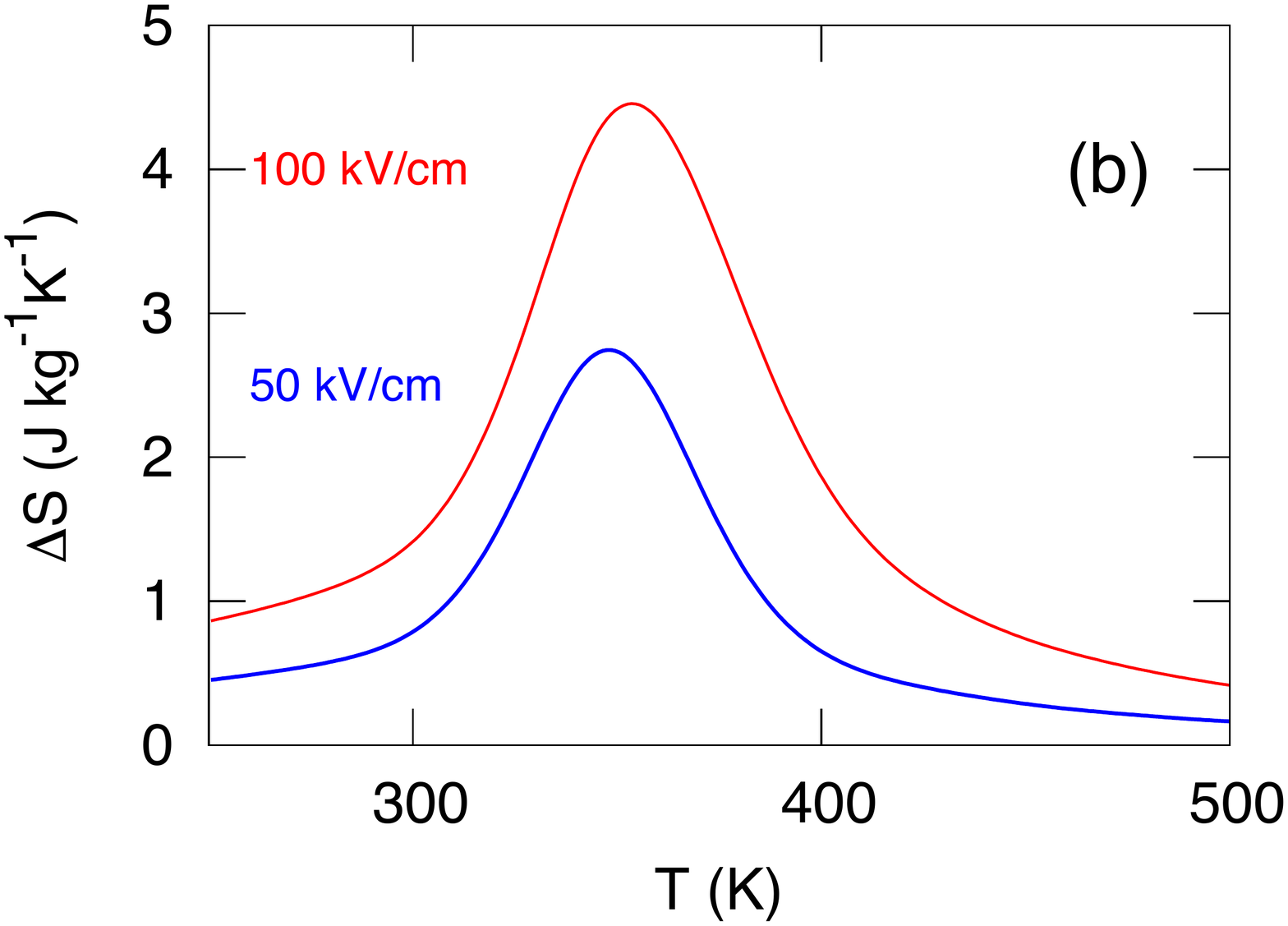}
\caption{(a) Comparison of the adiabatic temperature change found by the direct (dotted lines) and indirect (solid lines) method for 1\% parallel defects with a strength corresponding to a soft mode amplitude of 0.2~{\AA}. Black: Field range: 0---100 kV/cm; Red: 0---50kV/cm. Lines with symbols: ramping down the field with a rate of 0.05kV/cm/fs otherwise instantaneous ramping has been used. (b) Corresponding entropy change found by the indirect method.  
 \label{fig:rampkonv}error bar
}
\end{figure}

First, in order to compare our calculated temperature changes with experimental results, one has to keep in mind that we treat only three degrees of freedom as dynamical variables during our MD simulations, see Sec.~\ref{sec:comp}.
 During the time evolution the algorithm thus takes the system from the state ($P_1$,$T_1$,${\cal{E}}_1$), to the new state ($P_2$,$T_2$,${\cal{E}}_2$) while the state for the actual number of degrees of freedom would be ($P_3$,$T_3$,${\cal{E}}_2$).
The leading error in the adiabatic temperature change can be corrected if we rescale our results with a factor of 15/3, i.e.\ by rescaling the number of the degrees of freedom.
 Due to the uncorrected error in the final polarization (P$_3$), an error smaller than $1$~K is induced in the calculated adiabatic temperature change, \cite{Nishimatsu3,Cp} which does not modify the trends of the caloric response for different kinds of defects.

Second, experimental ramping rates are not accessible in our simulations due to the computational effort. For pure BaTiO$_3$, extensive convergency tests with respect to the ramping rate have been performed in Ref.~\onlinecite{Cp}, showing full convergency for a ramping rate of 0.002kV/cm/fs. In addition, an errorbar of about  0.3~K has been found for the  ramping rate of 0.05 kV/cm/fs which is used in the present investigation.
Indeed, the adiabatic temperature change of BTO as obtained by the direct method with these ramping rates and the adiabatic temperature change as obtained by the indirect method coincide as soon as ${\cal E}_1$ and ${\cal E}_2$ are beyond the critical field strength, cf.\ Ref.~\onlinecite{Cp}.
Thus, the reliability and accuracy of the used direct method is fully supported by the comparison with the indirect one.
Furthermore, our calculated temperature change is in good qualitative agreement to experimental values, e.g. $\Delta T=$0.9~K has been found for an applied field of 12 kV/cm\cite{Moya2}
while we yield $\Delta T=$3.8~K for an applied field of 100 kV/cm.

In order to test the influence of defects on the accuracy of the direct approach, the rescaling of the degrees of freedom, and the fast ramping, we have performed a set of calculations  for 1\% polar defects parallel to the external field with a strength corresponding to 0.2~{\AA}, see Fig.~\ref{fig:rampkonv}~(a).
First, we have calculated the adiabatic response by ramping down a field from ${\cal E}_1=100$~kV/cm to ${\cal E}_2=0$~kV/cm with a field rate of 0.05 kV/cm/fs.
Second, we have switched off the field instantaneously. Third, the indirect method has been applied, using the high-temperature value of  $C_{p,{\cal E}}$=15 k$_B=$3.3~J/Kcm$^3$ as quantum mechanical effects are neglected within our simulations.
We note that the indirect method can be applied even for  ${\cal E}_2=$0~kV/cm as the  system is beyond its critical point due to the strong internal field induced by defect dipoles, see Sec.~\ref{sec:dipol}.

\begin{figure}
\includegraphics[width=0.5\textwidth,clip,trim=2cm 2cm 0cm 2cm]{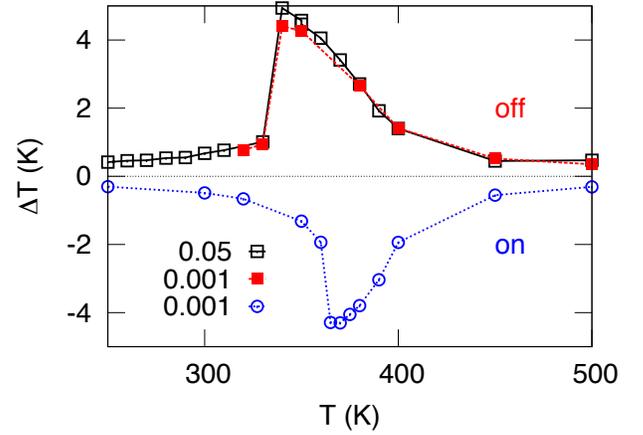}
\caption{(Color online) Dependency of the adiabatic response on the ramping rate for 1\% antiparallel defects with a strength corresponding to a soft mode amplitude of 0.4~{\AA}. The ramping rate is given in kV/cm/fs. Black and red: field removal, Blue: field on. 
 \label{fig:rampkonv2}
}
\end{figure}
Although, instantaneous switching may indeed bring the system out of equilibrium, see the detailed discussion in Ref.~\onlinecite{Cp}, the quantitative change of the caloric response compared to field ramping is already well below (below 0.25 K) the different trends of the adiabatic response discussed throughout this paper.
In addition, there is a rather good qualitative and semi-quantitative agreement between direct and indirect approach.
Small discrepancies can be understood by means of the temperature dependence of the specific heat. While the specific heat is approximately constant far from the phase transitions, there are pronounced peaks around $T_C$. Thus, especially at the maximum of the $\Delta T$ peak and at slightly lower temperatures, $\Delta T$ as obtained with the indirect method is too large due to the approximation used for $C_{P,\varepsilon}$. For example, if the specific heat calculated for the ideal system for an intermediate field strength, such as 75 kV/cm is used, the peak maxima between direct and indirect approach coincide, see the detailed discussion in Ref.~\onlinecite{Cp} \\
In summary, both direct and indirect method yield the same results, despite the different approximates made, underlining the good qualitative estimation of the temperature change found in our simulations and the small influence of possible errors discussed above.

In addition to the adiabatic temperature change ($\Delta T$), the isothermal entropy change ($\Delta S$) can be used to quantify the ECE.
In the indirect approach this quantity is given by the Maxwell relation as
\begin{equation}
\Delta S=\int_{{\cal{E}}_1}^{{\cal{E}}_2}\left(\left.\frac{\partial P}{\partial T}\right)\right|_{{\cal E}} d {\cal E}\;.
\label{eq:S}
\end{equation}
Figure~\ref{fig:rampkonv}~(b) exemplary shows the isothermal entropy change if an external field is removed from the sample with 1\% parallel defects.
For ${\cal{E}}_1=100$~kV/cm and ${\cal{E}}_2=0$~kV/cm we find a maximal increase in entropy of $\Delta S$=4.5~J$\cdot$kg$^{-1}$K$^{-1}$. This value is comparable to $\Delta S=7$~J$\cdot$kg$^{-1}$K$^{-1}$ found for ideal BTO and a field range from
${\cal{E}}_1=300$ kV/cm to ${\cal{E}}_2=75$ kV/cm.\cite{Cp}
It has to be noted that  $\Delta S$ and  $\Delta T$ are approximately related to each other by 
\begin{equation}
\Delta T \sim -\frac{T}{C_{P,\varepsilon}} \Delta S \,.\cite{Moya} 
\label{eq:S2}
\end{equation}
Thus, the same trends of $\Delta S$ and $\Delta T$ have to be expected for the various defect configurations discussed throughout this paper.

For antiparallel defects, the use of the indirect methods may fail, as the phase diagram depends on the history of the sample, e.g. cooling- down or heating-up simulations, see Fig.~\ref{fig:P_down_field}.
Instead, the convergency of the adiabatic temperature change with respect to the ramping rate is illustrated for antiparallel defects in Fig.~\ref{fig:rampkonv2}.
 1\% dipoles corresponding to 0.4~{\AA} have been used and an external field of 100 kV/cm has been removed or applied with two different ramping rates.
Obviously, the adiabatic temperature change down to starting temperatures of 300 K is sufficiently converged with respect to the ramping rate and a reversible temperature change is found for ramping the field on and off.
Analogous, the adiabatic temperature change for 1\% defect dipoles corresponding to 0.2~{\AA} as obtained with ramping rates of 0.5 or 0.001 kV/cm/fs differ less than 0.1~K for temperatures above 300~K.

We note, that for the different metastable states, which occasionally appear at lower temperatures or different combinations of internal and external field strength, the direction and ramping rate of the field may play a role. For example, for 1\% defect dipoles corresponding to 0.2~{\AA} $\Delta T$ as obtained with different ramping rates below 300~K differs by 0.5~K.
We leave a closer inspection of this parameter range for upcoming studies.

\bibliography{../../Diss/anna}

\end{document}